%% file: kcut.tex
\documentclass[11pt]{article}
\usepackage{times}
\usepackage{fullpage}
\usepackage{graphics}
\usepackage{epic}
\usepackage{psfig}
\usepackage{doublespace}

\newcommand{\nonest}[1]{\nestOff #1 \nestOn}
\newcommand{\nestOn}{\renewcommand{\nonest}[1]{\nestOff ##1 \nestOn}}
\newcommand{\nestOff}{\renewcommand{\nonest}[1]{}}

\newif\ifdraft
\draftfalse

\ifdraft
\newcommand{\markchange}[4]{%
  $^{\mbox{\tiny #1}}${#2}$^{\mbox{\tiny #1}}$%
  \nonest{%
    \footnote{{\em Was} ``{\small #3}''. {\em #1: #4}}%
    \marginpar{#1\arabic{footnote}}}}
\newcommand{\comm}[2]{\begin{quote}\marginpar{#1}{\tt #2}\end{quote}}
\else
\newcommand{\markchange}[4]{#2}
\newcommand{\comm}[2]{}
\fi

\newcommand{\ny}[3]{\markchange{NY}{#1}{#2}{#3}}
\newcommand{\dk}[3]{\markchange{DK}{#1}{#2}{#3}}
\newcommand{\newmt}[3]{\markchange{MTnew}{#1}{#2}{#3}}
\newcommand{\mtcom}[1]{\comm{MT}{#1}}

\long\def\comment#1{}
\def\etal{et al.\ }

\newcommand{\set}[1]{\{#1\}}
\newcommand{\R}{{\sf R\hspace*{-0.9ex}\rule{0.15ex}{1.5ex}\hspace*{0.9ex}}}

\newcommand{\vol}{\mbox{vol}}

\newtheorem{theorem}{Theorem}[section]

\newtheorem{lemma}[theorem]{Lemma}
\newtheorem{corollary}[theorem]{Corollary}

\newtheorem{fact}[theorem]{Fact}

\newenvironment{linearprogram}[2]{

\begin{eqnarray*}
& & \mbox{#1\ } #2 \\
& & \mbox{subject to\ }\left\{ % } MATCHING
\begin{array}{rcl@{\hspace{0.2in}}l}}{
\end{array}\right.
\end{eqnarray*}
}

\newcommand{\qed}{\vspace{.1em}\noindent\fbox{\rule{%
0em}{.1em}\rule{.1em}{0em}}\vspace{1em}}

\newenvironment{proof}{%

\noindent{\bf Proof:}\ }{%
\hfill \qed

}

\begin{document}

\title{Rounding Algorithms for a Geometric Embedding of Minimum Multiway Cut}
\author{%
  David R. Karger\thanks{\sloppy\scriptsize MIT Laboratory for
    Computer Science, 
    Cambridge, MA 02138. {\tt karger@lcs.mit.edu.} Research supported
    by NSF contract CCR-9624239, an 
    Alfred P. Sloane Foundation Fellowship, and a David and Lucille
    Packard Foundation Fellowship.
    }
  \and
  Philip Klein\thanks{\scriptsize Brown University . {\tt
      klein@cs.brown.edu.} 
    Research supported by NSF Grant CCR-9700146.}
  \and
  Cliff Stein\thanks{\scriptsize
    Dartmouth College.
    {\tt cliff@cs.dartmouth.edu.}
    Research supported by NSF Career award CCR-9624828.
    }
  \and
  Mikkel Thorup\thanks{\scriptsize AT\&T Labs--Research, Shannon Laboratory,
    180 Park Avenue, Florham Park, NJ 07932.
    {\tt mthorup@research.att.com.}}
  \and
  Neal E. Young\thanks{\scriptsize
     Akamai Technologies.
    {\tt neal@young.name}
    Research supported by NSF Career award CCR-9720664.
    }
  }

\date{\small\today}
\maketitle

\begin{abstract}
  {\small Given an undirected graph with edge costs and a subset of
    $k\ge 3$ nodes called {\em terminals}, a multiway, or $k$-way, cut
    is a subset of the edges whose removal disconnects each terminal
    from the others.  The multiway cut problem is to find a
    minimum-cost multiway cut. This problem is Max-SNP hard.  Recently
    Calinescu, Karloff, and Rabani (STOC'98) gave a novel geometric
    relaxation of the problem and a rounding scheme that produced a
    $(3/2-1/k)$-approximation algorithm.
  
    In this paper, we study their geometric relaxation.  In
    particular, we study the worst-case ratio between the value of the
    relaxation and the value of the minimum multicut (the so-called
    integrality gap of the relaxation).  For $k=3$, we show the
    integrality gap is $12/11$, giving tight upper and lower bounds.
    That is, we exhibit a family of graphs with integrality gaps
    arbitrarily close to $12/11$ and give an algorithm that finds a
    cut of value $12/11$ times the relaxation value.  Our lower bound
    shows that this is the best possible performance guarantee for any
    algorithm based purely on the value of the relaxation.  Our upper
    bound meets the lower bound and improves the factor of 7/6 shown by
    Calinescu et al.
    
    \ny{For all $k$, we show that there exists a rounding
      scheme with performance ratio equal to the integrality
      gap, and we give explicit constructions of polynomial-time rounding
      schemes that lead to improved upper bounds.}{We also
      improve the upper bounds for all larger values of $k$.}{}
    For $k=4$ and $5$, our best upper bounds are based on
    computer-constructed rounding schemes (with computer
    proofs of correctness).  For \ny{general $k$}{$k>6$}{} we give an algorithm
    with performance ratio $1.3438-\epsilon_k$.

    Our results were discovered with the help of computational
    experiments that we also describe here.  
}
\end{abstract}

\section{Introduction}
\label{sec:intro}
As the field of approximation algorithms matures, methodologies are
emerging that apply broadly to many NP-hard optimization problems.
One such approach (c.f.~\cite{LeightonRao88,LLR,AR,GW,EvenNRS}) has
been the use of metric and geometric embeddings in addressing graph
optimization problems.  Faced with a discrete graph optimization
problem, one formulates a relaxation that maps each graph node into a
metric or geometric space, which in turn induces lengths on the
graph's edges.  One solves this relaxation optimally and then derives
from the relaxed solution a near-optimal solution to the original
problem.

%oldney wording:
\iffalse
Recently Calinescu, Karloff, and Rabani followed the approach introducing
a new geometric relaxation for the min-cost multiway cut problem 
\cite{CKR98}. Their solution to the geometric relaxation did provide
strongly improved performance guarantees for the multiway cut problem. 
However, their solution to the geometric relaxations was not obviously
optimal, and indeed, in this paper, we provide improved solutions to the
geometric relaxation, thus further improving the approximation guarantees for
the multi-way cut problem. 

\paragraph{Multiway Cut.} 
The {\em min-cost multiway (or $k$-way) cut problem\/} is the following
\fi

This approach has been applied successfully~\cite{CKR98} to the {\em
  min-cost multiway cut problem}, a natural generalization of the
  minimum $(s,t)$-cut problem to more than two terminals.  An instance
  consists of a graph with edge-costs and a set of distinguished nodes
  (the {\em terminals\/}).  The goal is to find a minimum-cost set of
  edges whose removal separates the terminals.  If the number of
  terminals is $k$, we call such a set of edges a $k$-way cut.

The first approximation algorithm for the multiway cut problem in
general graphs was given by Dahlhous, Johnson, Papadimitriou, Seymour,
and Yannakakis~\cite{Dahlhous JACM paper}.  It used a traditional
minimum $(s,t)$-cut algorithm as a subroutine and had a performance
guarantee of $2-2/k$.

In the work that prompted ours, Calinescu, Karloff, and
Rabani~\cite{CKR98} used a novel geometric relaxation of $k$-way cut
in a $(3/2-1/k)$-approximation algorithm.  Their relaxation uses the
{\em $k$-simplex\/} $\Delta=\{x\in\R^k : x \geq 0,\linebreak[2] \sum_i
x_i =1\}$, which has $k$ vertices; the $i^{th}$ vertex is the point
$x$ in $\Delta$ with $x_i=1$ and all other coordinates 0.  The
relaxation is as follows: map the nodes of the graph to points in
$\Delta$ such that terminal $i$ is mapped to the $i^{th}$ vertex of
$\Delta$.  Each edge is mapped to the straight line between its
endpoints.  The goal is to minimize the {\em volume\/} of $G$,
$$  \mbox{vol}(G) \doteq \sum_{\mbox{\scriptsize edges } e}
\mbox{cost}(e) \cdot |e|
$$ where cost$(e)$ is taken to be the cross-sectional area of edge $e$
and $|e|$ denotes the {\em length\/} of the embedded edge $e$, defined
as half the $L_1$ distance between its endpoints.  The problem of finding an embedding that minimizes the volume can be formulated as a linear program (LP).
The factor half in
the length function is present to scale the distance between
terminals to 1, so the LP is a relaxation
of the minimum $k$-way cut problem.
%%%KARGER In the introduction, we talk say "ensure that the LP is a relaxation
%%%of the minimum k-way  cut problem".  This is a problem because we have
%%%not yet said anything at all about using a linear program.  Should
%%%probably talk about the embedding being a relaxation.

To see that the given LP is a relaxation of $k$-way cut, consider any
$k$-way cut and let $S_i$ be the set of nodes reachable from terminal
$i$ in the graph with the cut-edges removed.  Consider a geometric
embedding in which all nodes in $S_i$ are mapped to vertex $i$ of
$\Delta$.  For any edge, its embedded length is either 0, if the
endpoints lie in the same $S_i$, or 1, if the endpoints lie in
distinct $S_i$.  Hence the volume of this embedding is equal to the
cost of the $k$-way cut.

The algorithm of Calinescu \etal finds a minimum volume embedding by
linear programming.  It then uses a randomized rounding scheme to
extract a cut from this embedding.  Ignoring the graph, the scheme
chooses (from a carefully selected distribution) a {\em $k$-way cut of
  the simplex\/}---a partition of the simplex into $k$
polytopes, each containing exactly one vertex of the
%%%KARGER: here we talk about k-cut of simplex being partition into
%%%polytopes; later we talk about partition into arbitrary subsets.
%%%Need to be consistent.
simplex.  The $k$-way cut of the simplex naturally induces a $k$-way
cut in the embedded graph---namely, the set of edges with endpoints in
different blocks of the partition.  This cut has expected cost at most
$3/2-1/k$ times the volume of the embedding.

\subsection{Our results}

Our goal is to further understand the geometric relaxation, with the
hope of developing better approximation algorithms.  We aim to
determine the {\em integrality gap\/} of the relaxation and to find an
algorithm whose approximation ratio matches the integrality
gap.   \ny{(Formally, the integrality gap is the
  supremum, over all weighted graphs $G$,
  of the minimum cost of any $k$-way cut of $G$ divided by
  the minimum volume of any embedding of $G$.}{}{}
Note that the integrality gap is the best approximation ratio we can
\ny{prove using an analysis that bounds the optimum cut
only by the value of the relaxation.)}{
achieve for an algorithm that compares itself only to the embedding
volume.}{}

In this paper, we resolve this question for $3$-cut and provide
improved results for the general $k$-cut problem.  For $k=3$ we give a
rounding algorithm with performance ratio $12/11$, improving the Calinescu
\etal bound of $3/2-1/3=7/6$.  We also show that $12/11$ is the
best possible bound, exhibiting a graph family with a gap of
$12/11-o(1)$ between its embedded volume and minimum $3$-way cut.
Thus, for $k=3$, we determine the exact integrality gap and give an
optimal rounding algorithm.

For larger $k$, we obtain results based on both
computation and analysis.  \ny{We give a non-constructive
  proof that, for every $k$, there exists a \dk{(not necessarily
    polynomial time)}{}{} rounding scheme
  whose performance guarantee equals the integrality gap.}{}{}
For $k=4,5$, we use LP-derived and
-analyzed rounding schemes to give explicit approximation bounds of $1.1539$ and $1.2161$
respectively, improving the corresponding  Calinescu \etal
bounds of $1.25$ and $1.3$.  For larger $k$ we give an algorithm
obtaining a (analytic) bound of $1.3438-\epsilon_k$ where
$\epsilon_k>0$.  The quantity $\epsilon_k$ can be evaluated
computationally for any fixed $k$; we use this to prove that
$1.3438-\epsilon_k < 3/2-1/k$ for all $k$.

Our efforts to find geometric cutting schemes that achieve good
guarantees were guided by experiments: we formulated the problem of
determining an optimal probability distribution on $k$-way cuts of the
simplex as an infinite-dimensional linear program and solved discrete
approximations of this linear program and its dual. From these
solutions we were able to deduce the lower bound and, using that, the
upper bound for $k=3$.  These experiments also guided our search for
cutting schemes that work for larger values of $k$.

The upper and lower bounds for $k=3$ were discovered independently by
Cunningham and Tang~\cite{CunninghamT99}.

\paragraph{Presentation overview.}

In Section~\ref{sec:prelim} we discuss the geometric ideas underlying
the problem.  In Section~\ref{sec:experiments} we describe the
computational experiments we undertook and the results it gave for
small $k$.  In Sections~\ref{sec:ub3} and~\ref{sec:lb3} we resolve the 
3-terminal case, giving matching upper and lower bounds.  Finally, in
Section~\ref{sec:ub}, we present our improved algorithm for general
$k$.  \ny{In the appendix we prove that, for all $k$,
  there exists a rounding scheme matching the integrality gap.}{}{}

\section{The geometric problem}
\label{sec:prelim}

Finding the integrality gap of and a rounding scheme for the
relaxation turns out to be expressible as a  
geometric question.  That is, we can express integrality gaps
and algorithmic performance purely in terms of the simplex, without
considering particular graphs or embeddings.

Consider an edge $e$, which under the relaxation is embedded as a line
segment in the simplex.  We overload $e$ to denote this embedded
segment as well. 
For any segment (or edge) $e$, we let $e_\ell$ denote the
  projection of $e$ onto the $\ell^{th}$ coordinate axis, namely the
  one dimensional interval $\{x_\ell \mid x \in e\}$. We write $\min
  e_\ell=\min_{x\in e}x_\ell$ for the minimum value in the projected
  interval, $\max e_\ell= \max_{x\in e}x_\ell$ for the maximum value
  in the projected interval, and $|e_\ell|=\max
  e_\ell-\min e_\ell$.
Finally, as mentioned in Section~\ref{sec:intro}, the length $|e|$ of an edge $e$ is defined to be half
its $L_1$ norm, that is,
\[
|e|=\sum_{\ell=1}^k|e_\ell|/2.
\]

\subsection{Density}
\label{sec:density}

Recall that a $k$-way cut of the simplex is a partition of the simplex
into $k$ polytopes, each containing a unique vertex of the simplex, and
that such a cut induces a $k$-way cut of any embedded graph.  By a
{\em cutting scheme}, we mean a probability distribution $P$ on
$k$-way cuts of the simplex.  For any line segment $e$, the {\em
  density of $P$ on segment $e$}, denoted $\tau_k(P,e)$, is the
expected number of times a random cut from $P$ cuts $e$, divided by
the length $|e|$ of $e$.\footnote{Note that in principle a line segment could be
  cut more than once by the $k$-way cut of the simplex.  We therefore
  speak of the expected number of times that $e$ is cut, rather than
  the probability that $e$ is cut.}

  Define the
{\em maximum density of $P$,} $\tau_k(P)$ and the {\em minimal maximum
  density} $\tau_k^*$ as follows:
\[ {\small
\tau_k(P) \doteq \sup_e \tau_k(P,e)
\mbox{~~~~and~~~~}
\tau_k^* = \inf_P \tau_k(P), }
\] 
There is always a line segment of infinitesimal length that achieves
the maximum density, since any segment can be divided into two 
edges, one of which has density no less than the original.  Thus, in
the remainder of this paper, we will focus discussion on such
infinitesimal segments.

The relevance of $\tau_k^*$ is the
following (this is implicit in the work of Calinescu et al.):
\begin{lemma}\label{densitylemma} 
  For any cutting scheme $P$ and embedded graph $G$, the expected cost
  of the $k$-way cut of $G$ induced by a random $k$-way cut from $P$
  is at most $\tau_k(P)$ times the cost of the embedding of $G$.
\end{lemma} 
\begin{corollary}
Any cutting scheme $P$ yields a randomized approximation algorithm with
  approximation ratio at most $\tau_k(P)$.
\end{corollary}
\begin{proof}
  The endpoints of any edge $e$ are embedded at two points in the
  simplex, so the edge corresponds to a segment connecting those two
  points.  The expected number of times the edge is cut
  is at most  $\tau_k(P,e)\cdot |e|$.  By the Markov inequality this upper
  bounds the probability that the edge is cut.  Thus, the expected
  cost of the $k$-way cut is at most
\begin{eqnarray*}
\sum_e (\tau_k(P,e)\cdot |e|)\mbox{cost}(e) 
&\le &\tau_k(P) \sum  |e|\cdot\mbox{cost}(e) \\
&=  &\tau_k(P) \vol(G).
\end{eqnarray*}
We have already argued that $\vol(G)$ lower bounds the optimum
$k$-way cut, so the result follows.
\end{proof}

\ny{
  The above argument implies that no cutting scheme $P$ can have a
  maximum density $\tau_k(P)$ below the integrality gap.
  In fact, we show that there always {\em exists} a cutting
  scheme whose maximum density {\em equals} the integrality gap.
  \begin{theorem}\label{infinite_case}
    There exists a cutting scheme whose maximum density equals
    the integrality gap, thus, $\tau^*_k$ equals the integrality gap.
  \end{theorem}
  We give the proof in the appendix.  The proof is based on
  the observation that the problem of choosing a rounding
  scheme to minimize the performance ratio is itself a (infinite dimensional) linear
  programming problem; furthermore its dual is the problem of
  choosing a weighted graph to maximize the integrality gap.
  This observation seems to hold in a fairly general setting
  beyond the $k$-cut problem (details are in the appendix).
}{}{}

Calinescu et al.'s algorithm gives a cutting scheme showing that $\tau_k^*
\le 3/2-1/k$.  In this paper we show that $\tau_3^* = 12/11$, and that,
for all $k$, $\tau_k^* \le 1.3438.$

\subsection{Alignment}\label{sec:align}

\newcommand{\plane}[2]{\Delta_{x_{{#1}}={#2}}}
\newcommand{\corner}[2]{\Delta_{x_{{#1}}\geq{#2}}} 

We have just argued that the key question to study is the maximum
density of (infinitesimal) line segments relative to a cutting scheme.
Calinescu \etal showed that one can restrict attention to segments in
certain orientations.  We say a segment $e$ in $\Delta$ is {\em
  $i,j$-aligned\/} if $e$ is parallel to the edge connecting vertices
$i$ and $j$ of $\Delta$.  We say it is {\em aligned\/} if it is
$i,j$-aligned for some pair of vertices.  Calinescu \etal observed
that the endpoints of any
segment $e$ can be connected by a piecewise linear path of total
length $|e|$ whose segments are aligned.  The segment $e$ is cut if and only if
some edge on this path is cut.  Given any embedding of a graph,
Calinescu \etal apply this transformation to
each segment connecting two embedded vertices, without changing the
volume of the embedding.  Thus, they show that without loss
of generality one may restrict attention to embeddings in which all
edges are aligned.

\begin{fact}  Segment $e$ is $i,j$-aligned if and only if
$|e|=|e_i|=|e_j|$ and $|e_\ell|=0$ for 
$\ell\neq i,j$.
\end{fact}
(Note that $|e|$ denotes half the $L_1$ norm, while $|e_1|$  and
$|e_2|$ are standard absolute values.)

\subsection{Side parallel cuts (SPARCS)}
\label{sec:sparc}

In this paper, we mainly restrict attention to a particular class of
cutting schemes.  Define $\plane i {\rho}\doteq\{x\in \Delta:x_i=\rho\}$ and
$\corner i {\rho}\doteq\{x\in \Delta:x_i\geq \rho\}$.  Note that
$\plane i {\rho}$ 
is a hyperplane that runs parallel to the
face of the simplex opposite terminal $i$ and is at distance $\rho$
from that face; 
it divides the simplex into two parts, of which $\corner i \rho$ is the
``corner'' containing terminal $i$.
An $i,j$-aligned segment $(x,y)$ is cut by the hyperplane
$\plane \ell \rho$ if and only if 
$\ell\in\{i,j\}$ and $\rho$ is between $x_\ell$ and $y_\ell$. 

We define a {\em
  side-parallel cut (sparc)} of the simplex:
\begin{enumerate}
\item Choose a permutation $\sigma$ of the vertices;
\item For each vertex $i$ in order by $\sigma$ (except possibly the last),
  choose some  $\rho_i\in[0,1]$;
\item \label{step:slice} Assign to vertex $i$ all points of
  $\corner{i}{\rho_i}$ not already assigned to a previous terminal.
  We say that terminal $i$ {\em captures} all these points, and that
  terminal $i$ {\em cuts\/} an edge $e$ if it captures some but
    not all of the yet-uncaptured part of $e$.
\end{enumerate}
This scheme cuts up the simplex using hyperplanes $\plane i {\rho}$.
In this context, we 
call each $\plane i {\rho}$ a {\em slice}.

We consider algorithms that sample randomly from some
probability distribution over sparcs.  Our restriction to
sparcs was motivated by several factors.  The rounding
algorithm of Calinescu \etal uses only sparcs.  Furthermore,
our computational study of the 3-terminal problem (discussed
below) and some related analytic work gave some evidence that
the optimal algorithm was a distribution over sparcs (this
  conjecture was confirmed analytically for the 3-terminal
  case).  Lastly, sparcs have concise descriptions (as
sequences of $k-1$ slicing distances) that made them easy to
work with computationally and analytically.  It is
conceivable, though, that one might do better with cuts that
are not sparcs.  For example, one might wish to slice off
  two terminals simultaneously, and then separate the two from
  each other.  Indeed, we know of no proof
  that for $k>3$ the optimal cut must be made up of
  hyperplanes; curved surfaces might do better.

%Note that Step~\ref{step:slice} can also be described as passing the
%hyperplane $\plane i {1-d_i}$ through the simplex, parallel to the face
%opposite terminal $i$ and at distance $d_i$ from terminal $i$, and
%assigning to terminal $i$ all points on the near side of the
%hyperplane.  To avoid confusion with cuts, we will refer to one such
%hyperplane of the cut as a {\em slice.}

For segment $e$, recall that $e_\ell$ is the interval $\set{x_{\ell} \mid
x\in e}$.  The key properties of sparcs are expressed in the
following fact.
\begin{fact}\label{sparc-prop} An $i,j$-aligned segment $e$ is cut by a sparc
  if and only if it is cut by terminal $i$ or $j$. Furthermore, for $\ell\in\{i,j\}$, 
the following conditions are all necessary  for segment $e$ to be cut 
by terminal $\ell$:
\begin{itemize}
\item[(1)] $\rho_\ell\in e_\ell$
\item[(2)] For all terminals $h$ preceding $\ell$, $\rho_h>\min e_h$.
\item[(3)] Terminal $\ell$ is not last in the order
\end{itemize}
\end{fact}

For the following, let $e$ be an $i, j$-aligned segment.  For
probability distributions $P$ on sparcs, one can obtain bounds on 
$\tau_k(P,e)$ by using Conditions~1--3 above.  For example, we can
restrict our attention to Condition~1: If $\rho_i$ and $\rho_j$ are
uniformly distributed over $[0,1]$, Condition~1 holds for terminal $i$
with probability $|e_i|=|e|$, and similarly for terminal $j$. Thus, by
linearity of expectation, the expected number of times $e$ is cut is
at most $2 |e|$.

Next, consider adding Condition~3.  Suppose that the
ordering of terminals is random, meaning that $i$ is last with
probability $1/k$.  The probability that $e$ is cut by $i$ becomes
$(1-1/k)|e|$, so $\tau_k(P,e) \leq (2-2/k)$. Thus, uniformly random
$\rho_{\ell}$'s and a random ordering gives a performance guarantee of
of $2-2/k$, matching the bound of Dahlhous et al.~\cite{Dahlhous JACM
  paper}.

To improve these bounds, one must use Condition 2.  Calinescu \etal
choose a sparc by selecting $\rho$ uniformly at random in $[0,1]$,
setting $\rho_{\ell} = \rho$ for each terminal $\ell$, and slicing off
terminals in random order.  Conditions 1 and 3 again derive a density
bound of $1-1/k$ for terminal $i$ and $j$.  Calinescu \etal  improve
this analysis as follows.  Suppose that the edge $e$ is farther from $j$
than from $i$.  We will argue that the density contribution from
terminal $j$ (i.e. the contribution to $\tau_k(P,e)$ due to terminal
$j$ cutting $e$) is only 1/2.  The point is that if $\rho$ is such that
$j$ potentially cuts $e$, i.e. $\rho \in e_j$, and if $i$ (which is
closer to $e$) precedes $j$ in the random 
slice ordering, then $i$ will capture all of $e$ and prevent $j$ from
cutting it.

Formally, we argue as follows.  Without loss of generality, assume
that $\min e_i\geq \max e_j$ (note that any $i,j$ aligned edge can be
split in two with one part closer to $i$ and one part closer to $j$,
and our assumption then applies to each part separately).  As was
  argued before, the contribution of terminal $i$ is at most
  $1-1/k$.  On the other hand, with probability 1/2,
$i$ precedes $j$. If so, since $\rho_i=\rho_j$, Condition 1---that
$\rho_j\in e_j$---contradicts Condition 2 for $i$---that $\rho_i>\min
e_i$. Thus $e$ can only be cut by terminal $j$ if $j$ precedes $i$, in
which case by Condition 1, the density contribution from $j$ is 1.
Thus the density contribution from terminal $j$ is 1/2,
leading to a total density of $3/2-1/k$.

To improve on the 3/2 bound, we made stronger use of Condition 2.  The
analysis of Calinescu \etal only considers that a segment may be
captured by the two terminals with which it is aligned.  We derive
stronger results by observing that other terminals may capture the
segment as well.  To do so, we had to change the cut distribution as
well as the analysis.  It can be shown that no distribution that holds
all $\rho_i$ equal can do better in the limit than the $3/2$ factor of
Calinescu \etal. For independent, uniformly distributed $\rho_i$
we also get the 3/2 factor. 
The 3/2 factor can be improved
somewhat in the limit by using a non-uniform distribution on each
  (independent) $\rho_i$. However, the best cutting
schemes we have found are based on combining dependence and
  non-uniformity.  One such scheme for 3-way cut gives us a bound of
$12/11$, which is optimal over all schemes for 3-way cut.  Another
scheme gives us a bound of $1.3438$ that holds for any number of
terminals.  This latter scheme is designed for large $k$;
  optimizing it for smaller $k$ gives better bounds.

%%% CCC replaced with one sentence above
\iffalse
Though they are natural, sparcs are not the only cuts one might
imagine using. 
 sparcs effectively cut one terminal off the simplex at
a time, and thus do not explore all possible cut topologies.  
The key observation of Calinescu et al., which led to there improved
approximation algorithm, was that slices do not operate in isolation.
Rather, for an $i$--$j$ aligned edge, if the slice for terminal $i$
precedes that for $j$, then it has some chance of "capturing" the
entire edge, preventing the slice for terminal $j$ from cutting the
edge.  However, the analysis of Calinescu \etal considered only the
two slices that may actually cut the edge.  We generalize their
approach, observing that the other slices may also capture the edge
before the $i$ or $j$ slices get a chance to cut the edge.  This lets
us prove better bounds.
\fi

\subsection{Additional Observations}

\paragraph{What is the best embedding?}

Perhaps the first natural question to ask is whether the embedding
chosen by Calinescu et al. is the best possible.  

\begin{lemma}
  Among all relaxations based on embeddings in the simplex that
  minimize some norm (without adding other constraints) the $L_1$ norm
  has the smallest possible integrality gap.
\end{lemma}
\begin{proof}
\newcommand{\normp}[1]{\|#1\|'}%
\newcommand{\normo}[1]{|#1|}%
We show that the $L_1$ norm maximizes the measured volume of any
embedded graph; thus it minimizes the integrality gap.

Suppose we use some norm $\normp{\cdot}$ instead of the the (scaled)
$L_1$ norm $\normo{\cdot}$.  If the norm provides a relaxation, the
distance between simplex vertices must be at most one\dk{---that is, for
any edge $e$ connecting simplex endpoints, we have $\normp{e} \le 1 =
\normo{e}$}{}{}.  Consider some 
embedded edge $e$.  As discussed in Section~\ref{sec:align}, we know
that under the $L_1$ norm it is connected by a path of
aligned edges $e_1,\ldots,e_r$ such that $\normo{e}=\sum\normo{e_i}$.
Since any norm-based distance measure is translation invariant and
proportion preserving, this implies that for each $e_i$ (which is a
scaled, translated version of an edge of the simplex) we have
$\normp{e_i} \le \normo{e_i}$.  It follows from the triangle
inequality that
\begin{eqnarray*}
\normp{e} &\le &\sum \normp{e_i}\\
&\le &\sum \normo{e_i}\\
&=&\normo{e}.
\end{eqnarray*}
Since $\normp{\cdot}$ assigns no greater a length to every embedded
edge than $\normo{\cdot}$,
it also assigns no greater a volume to any
embedded graph.  Thus, its integrality gap is no better than that
induced by the $L_1$ norm.
\end{proof}

\paragraph{Symmetry.}

A second observation is that there is no benefit in
trying to identify a ``good terminal order'' in which to cut up the
simplex.

\begin{lemma}
There is a sparc cutting scheme whose maximum density is minimum among
all sparc cutting schemes and that has the following form:
\begin{enumerate}
\item choose slice distances $(d_1,\ldots,d_{k-1})$ from some
  probability distribution 
\item apply the slice distances (in order) to a uniform random
  permutation of the terminals
\end{enumerate}
\label{symmetry-lemma}
\end{lemma}
An analogous ``order independence'' statement holds for the best
possible (possibly non-sparc) algorithm.
\begin{proof}
For any cutting scheme $P$, let $P'$ the the corresponding
``symmetrized'' cutting scheme, i.e.
\begin{eqnarray*}
\lefteqn{\Pr[P' \mbox{ cuts corner 1 at distance $d_1$, then cuts corner 2 at
    distance $d_2$, etc.}]} \\
& = \frac{1}{k!} \sum_{\sigma} \Pr[P\ \mbox{cuts corner $\sigma(1)$ at
    distance $d_1$, then\ldots}]
\end{eqnarray*}
where $\sigma$ varies over all permutations of $1, \ldots, k$.

For a line segment $e$ and a permutation $\sigma$, let $\sigma(e)$
denote the line segment obtained by permuting by $\sigma$ the
coordinates of the start point and end point of $e$.

Then for any line segment $e$,
\begin{eqnarray*}
\mbox{density of $P'$ on } e & = & \frac{1}{k!} \sum_{\sigma} \mbox{
  density of $P$ on } \sigma(e)
\end{eqnarray*}

Let $\tau$ be the maximum density of $P$.  Then for any $\sigma$, the
density of $P$ on $\sigma(e)$ is at most $\tau$.  The density of $P'$
on $e$ is thus the convex combination of values all of which are at
most $\tau$, so is in turn at most $\tau$.  Thus the density of $P'$
is no more than that of $P$.
\end{proof}

The above lemma shows that there is no worst-case benefit to
considering specific terminal ordering.  The duality argument of
Section~\ref{sec:density} carries over to show that a sparc with
optimum expected maximum density can be specified simply as a
distribution over slicing distances, without reference to an input
graph embedding.

\section{Our Computational Study}
\label{sec:experiments}
\iffalse
\begin{quote}
{\bf Neal said:} The section on "LP results" disregards the initial
work done with Maple and 
LP\_SOLVE (not cplex).  I don't know if we care (is crediting the software
important?)  

We've overlooked the point that the dual solutions were much more useful for
the k=3 case then the primal solutions.  Further, we do not describe the LOWER
BOUND LP, which was actually the most useful for k=3, and was not the same as
the upper bound LP that is described now.  

This is probably a subtle point not worth pursuing, but there is a more
important related issue.  With k=3 we were able to extrapolate the general
solution to the dual, but not to the primal, from the computational results.
Then we figured out an analytic primal solution by reasoning about what it had
to be like (according to complimentary slackness) if the analytic dual solution
(lower bound) was to be tight.  Recall we saw lots of optimal but very
different primal solutions (from our computations), whereas the dual solutions
tended to have a more canonical form.  Maybe this is particular to k=3, but
maybe not.  I think using a lower bound LP (or at least the dual) might be
important for higher k as well.

In fact I think we should pursue this lead ourselves by thinking carefully
about the dual that we obtain when we restrict to sparc cuts.

{\bf Mikkel:} I don't think we need to present this to the reader, so
I wouldn't worry. Looking at the duals ourselves could be interesting, but
first of all, I'd like this project to be completed and the journal version 
submitted.
\end{quote}
\fi

In this section we describe some computational experiments we carried
out to help us understand the behavior of the geometric embedding.
These experiments also yielded the best rounding schemes so far for
the 4- and 5- terminal cut problems.  One need not read this section in order to
understand the following ones.

As discussed above, our goal was to find a distribution over cuts of
the $k$-simplex that minimized the density of any  segment in
the simplex.  This problem can be formulated as an infinite
dimensional linear program, with one variable per cut of the simplex,
corresponding to the probability that that cut is chosen, and one
constraint for every (aligned, infinitesimally small) line segment
inside the simplex, which measures the expected number of times the
chosen cut will cut that segment.  Of course, it is not tractable to
solve the infinite LP computationally, but we expected that
discretized versions of it would be informative.

We applied this approach in two distinct ways.  For the 3-terminal
case, we devised an LP that exploited the planarity of the 3-terminal
relaxation to home in on a ``worst case'' embedded graph.  By
examining this graph, we were able to deduce requirements for the
optimal algorithm, which led to its identification.  For the general
case, we devised an LP whose solutions are (provable) upper bounds on
the performance of certain rounding algorithms.  We solved this LP for
small numbers of terminals (3--9), deriving algorithms with (computer
aided) proofs of the best known performance ratios for these problems.
The solution suggested certain properties that appear to hold in the
``optimal'' rounding scheme; we used these suggestions in our
development of (analytic) solutions for arbitrary numbers of
terminals.

\subsection{The three-terminal case}

\begin{figure}
\centerline{\psfig{figure=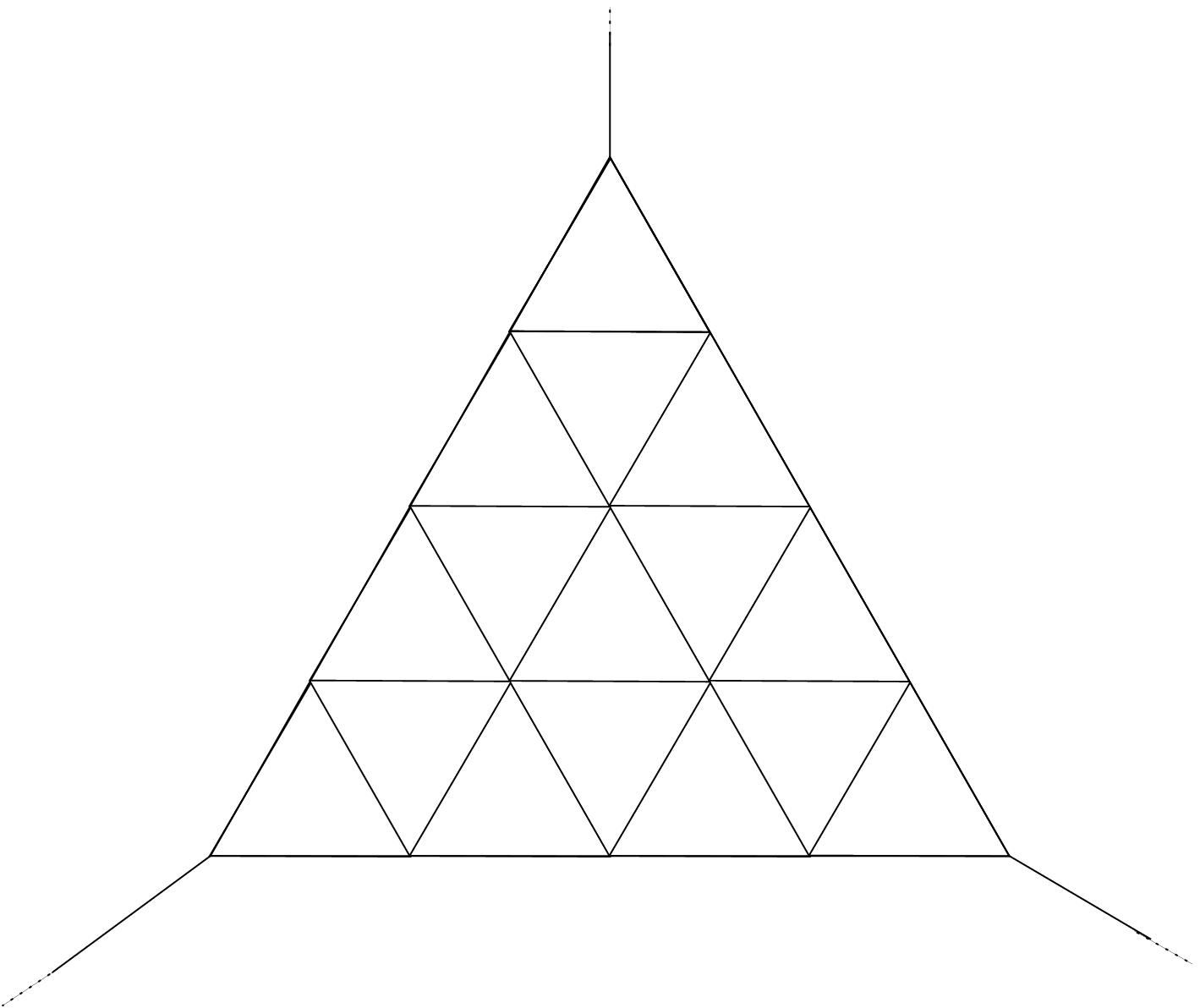,width=2.5in}}
\caption{A triangular mesh used in the linear program for the
  three-terminal case.  The mesh is augmented with rays going from the
  corners of the simplex to infinity.}
\label{triangular-grid-fig}
\end{figure}

For the $3$-terminal problem we exploited planarity.  The
3-simplex can be viewed as a triangle in the plane.  We discretized
the linear program by defining a triangular mesh over the simplex and
considering only edges of the mesh instead of all line segments in the
simplex. A (rather coarse) example mesh is shown in
Figure~\ref{triangular-grid-fig}.  Note that we have augmented the
triangular mesh with rays starting at the corners of the simplex and
heading out to infinity.

\begin{figure}
\centerline{\psfig{figure=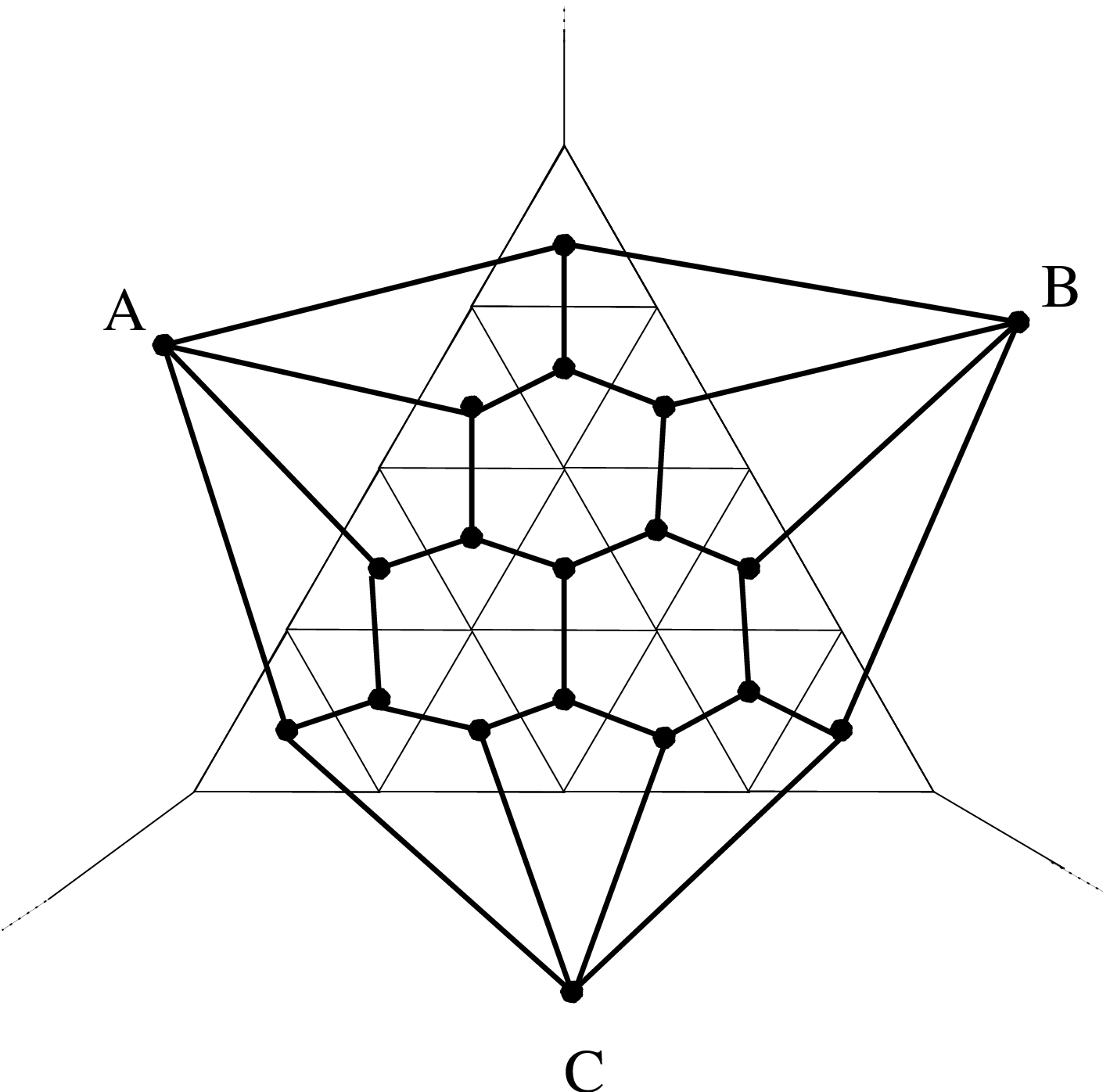,width=3in}}
\caption{The planar dual is shown in bold.  The nodes $A$, $B$, and
  $C$ correspond to the three infinite regions of the primal.}
\label{triangular-grid-with-dual-fig}
\end{figure}

We used the planarity of
the 3-simplex to simplify our LP formulation.  The planar dual of the
augmented mesh is shown in
Figure~\ref{triangular-grid-with-dual-fig}.  Note that because of the
augmentation, the dual has three {\em auxiliary nodes} $A$, $B$, and $C$ corresponding to infinite
regions of the primal, one node for each side of the simplex.

\begin{figure}
\centerline{\psfig{figure=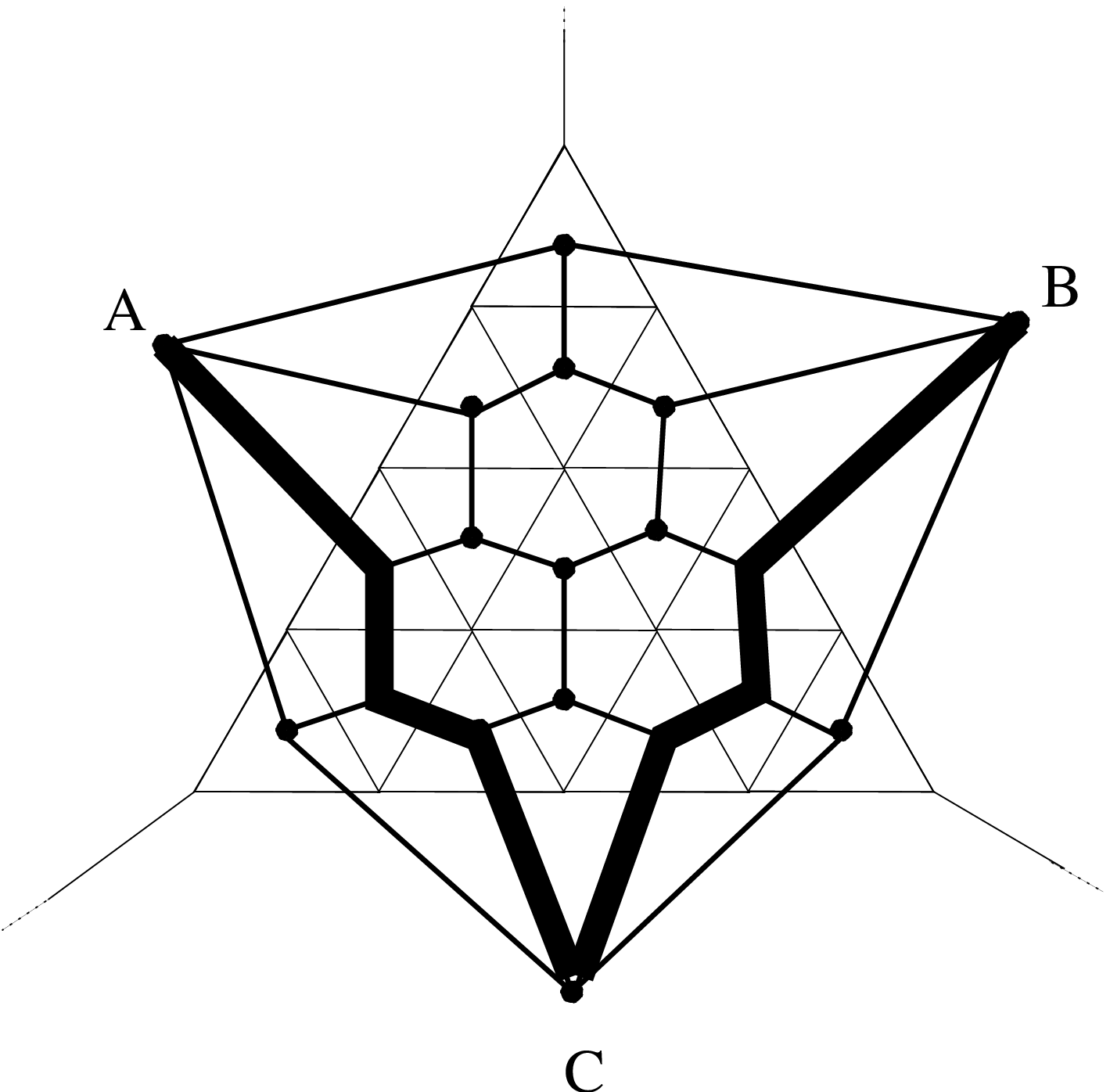,width=3in}}
\caption{An example of a 3-way cut corresponding to a pair of paths.}
\label{two-paths-fig}
\end{figure}

\begin{figure}
\centerline{\psfig{figure=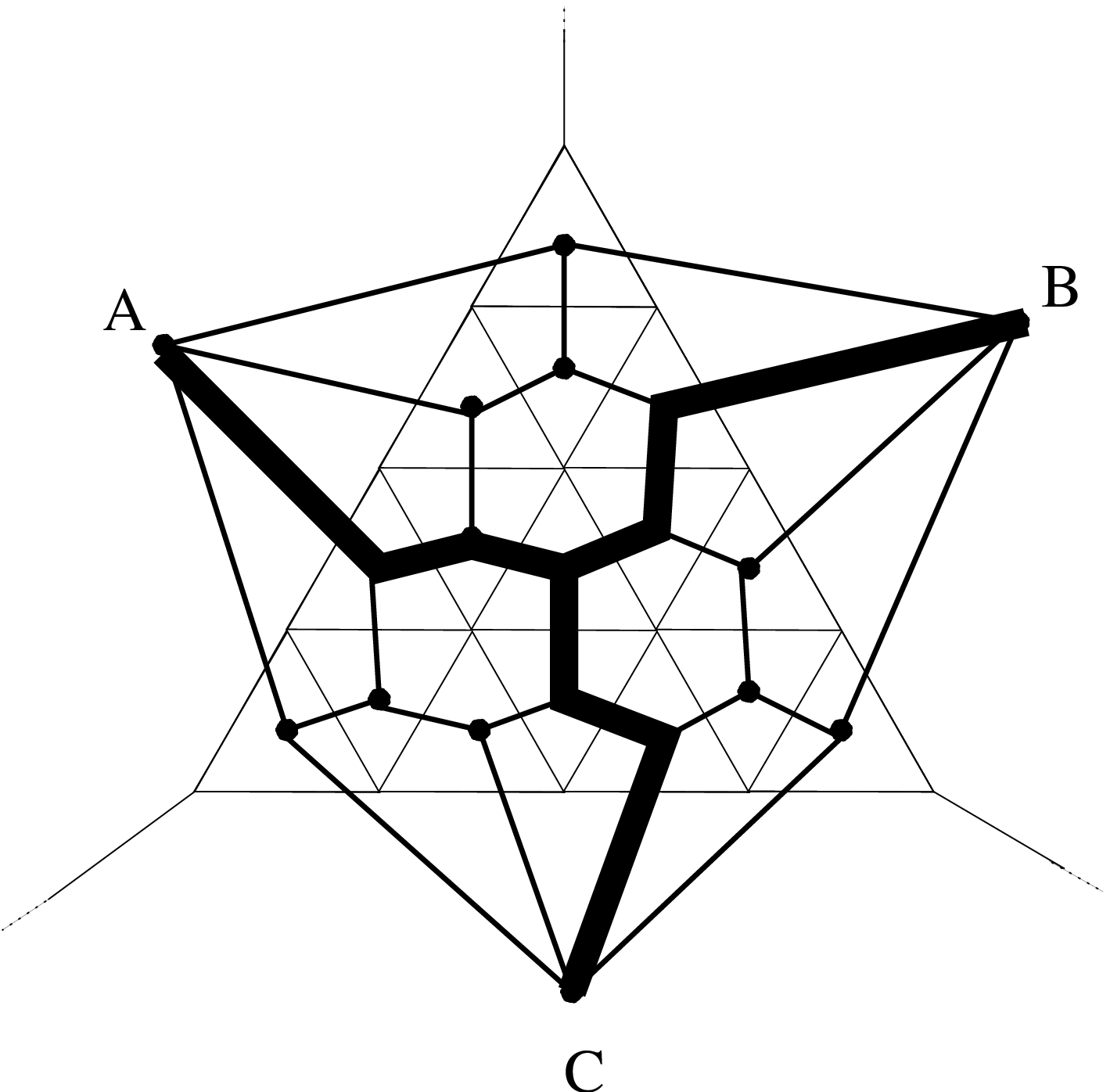,width=3in}}
\caption{An example of a 3-way cut corresponding to three paths.}
\label{three-paths-fig}
\end{figure}

Any minimal 3-way cut
of the mesh corresponds to a collection of two or three paths
(representing the boundary of the cut) through the planar dual of the
augmented mesh.  Specifically, the cut corresponds to either
\begin{enumerate}
\item two paths \label{two-paths}
whose endpoints are the nodes $A$, $B$, and $C$ (illustrated in Figure~\ref{two-paths-fig}), or 
\item three paths all originating at some interior dual node and
terminating at the nodes $A$, $B$, and $C$ (illustrated in
Figure~\ref{three-paths-fig}).\label{three-paths}
\end{enumerate}
Given an assignment of weights to the edges of the mesh, the weight of
the minimum 3-way cut of the first type is
$$\min_{X \in \set{A, B, C}} \mbox{sum of distances in dual graph from
  $X$ to each of the other auxiliary nodes}
$$
and the minimum 3-way cut of the second type is
$$\min_{x \mbox{\small\ ordinary node of dual}} \mbox{sum of distances in dual graph
  from $x$ to $A$, $B$, and $C$}$$

To find an embedded graph that is ``worst-case'' (up to the
discretization), we solve a linear program that has a weight variable $w_{uv}$
for each edge $uv$ of the triangular mesh (not including the rays).  The
linear program also has a distance variable $d_{xy}$ for every pair
$x,y$ of nodes of the dual (including auxiliary nodes).  The
objective is to minimize the total weight $\sum_e x_e$ subject to the
condition that every 3-way cut has value at least 1.  This condition
can be expressed by a collection of constraints on distances through
the dual graph.

\begin{eqnarray*}
\min \sum_{uv} w_{uv}
& s.t.
\end{eqnarray*}
\begin{eqnarray*}
d_{xA} + d_{xB} + d_{xC} \leq 1 & & \mbox{ for each ordinary dual node
}x\\
d_{AB} + d_{AC} \geq 1\\
d_{BA} + d_{BC} \geq 1\\
d_{CA} + d_{CB} \geq 1\\
d_{xz} \leq d_{xy} + w_{yz} & \mbox{ for each ordinary dual edge } yz\\
d_{xx} = 0 & \mbox{ for each dual node } x\\
w_{uv} \geq 0 & \mbox{ for each edge } uv\\
d_{xy} \geq 0 & \mbox{ for each dual edge } xy
\end{eqnarray*}

\iffalse REPLACED BY TEXT ABOVE --pnk, 2003
Thus the distribution of cuts corresponds to a packing of these
paths, which can be seen as a kind of flow.  So instead of enumerating
all possible cuts, we could define a linear program that assigned a
(multicommodity) flow to each edge of the dual mesh.  This gave us a
tractable representation of the linear program.

We also found it helpful to solve the dual of our flow-based linear
program, which assigns weights to the mesh edges to minimize the total
weight such that every $3$-way cut has value at least 1.  Since each
$3$-way cut corresponds to a set of two or three paths in the planar dual
of the mesh, the latter constraint can be represented efficiently by
constraining shortest-path lengths, which is to say distances (as a
function of the variable edge lengths) in the planar dual.  A solution
to the dual can be interpreted as a probability distribution over mesh
edges, which in turn can be interpreted as an embedded graph
demonstrating the integrality gap.  The dual solutions were more
informative than the primal solutions --- there seemed to be a larger
space of primal solutions than dual solutions, making it harder to pin
down the pattern. 
\fi %END OF REPLACED TEXT

Using the above linear program, we first deduced the general form of the dual
solution, giving us the lower bound for $k=3$. From this we deduced
the necessary structure of any optimal primal solution (using
complementary slackness conditions), including the important idea of
``ball cuts'' versus ``corner cuts'' which we will discuss in the
following sections.

\subsection{The general case}

In the general case, the lack of a planar embedding prevented us from
exploiting nice properties of its cuts; we were faced with the problem
of enumerating cuts as well as edges.  Based on the work of Calinescu
\etal and our own results for the optimal 3-terminal solution, we
decided to limit our exploration to sparcs as discussed above.

There is still an infinite space of possible sparcs, so we discretized
our problem.  Fix an integer {\em grid size} $N$.  A {\em discrete
  sparc} is described by a vector $(q_1,\ldots,q_{k-1})$ where each
$q_i$ is an integer in the range $[0,N-1]$.  Given such a vector, we
choose a random sparc by setting $d_i$ uniformly in the range
\mbox{$[q_i/N,(q_{i}+1)/N]$}.  This defines a probability distribution
on sparcs.  We now define a linear program to search for a probability
distribution over all discrete sparcs (which induces a probability
distribution over all sparcs).  We define a variable for each discrete
sparc, which reflects the probability of choosing that discrete sparc,
and provide constraints that upper bound the density of every
  possible segment under this probability distribution.  We then aim
  to minimize the largest of these densities.

There are infinitely many segments, but we define a finite set of
constraints that allow us to upper bound the density of all of them,
as follows. The slices at distances $q/N$ ($q = 1, 2, \ldots, N-1$) for each terminal partition the simplex into {\em
cells}
$$\set{(x_1, \ldots, x_k)\ :\ q_i/N \leq x_i \leq (z_i+1)/N}$$
For a given distribution on the discrete sparcs, we can
compute a (linear) upper bound on the density induced on {\em any}
segment with a given alignment within a cell, and specify one
constraint saying that this upper bound should be small.  Since the
cells are small, we expect all segments with a given alignment to have
roughly the same density under our cutting scheme, so we hope that the
upper bound is reasonably tight.  With this simplification, the number
of constraints is bounded by the number of cells times the number of
segment alignments per cell, which is at most $k^2 N^k$.

We determine the upper bound for a cell as follows.  For any discrete
cut, the slices generated from it will fall into one of three
categories.  If the $i^{th}$ coordinate of the discrete cut is
different from that of the cell, then the $i^{th}$ slice will not pass
through that cell: depending on whether the coordinate is larger or
smaller it will either capture the entire cell or none of the cell.
If the $i^{th}$ coordinates are the same, then the slice might pass
through the cell; we can use the fact that the slice is uniformly
distributed over a range to determine its density contribution.

An $i,j$-aligned segment can only be cut if the slices for terminal
$i$ or $j$ go through its cell (and no earlier slice captures the
  entire cell).  If only one of the two slices goes through the
terminal then its contribution to a segment's density is at most
$N$ (the length of the segment divided by the width of the
  cell). If both slices go through the cell, their
contribution is at most $2N$.  We ignore the fact that
different slices within the cell might capture the segment before it
can be cut, thus introducing some slack in our upper bound.

In view of Lemma~\ref{symmetry-lemma}, we
need only represent cutting schemes that consider terminals in random order.
Recall that each assignment of integers in $[0, N-1]$ to $q_1, \ldots, q_{k-1}$
defines a discrete sparc.  
We let $x_{q_1, \ldots, q_{k-1}}$ be the
variable assigning a probability to this sparc in the cutting scheme.
For any permutation $\sigma$ of $\set{1, \ldots, k}$, the probability
of applying this sparc to the sequence of terminals $\sigma(1),
\ldots, \sigma(k)$ is $\frac{1}{k!} x_{q_1, \ldots, q_{k-1}}$.

In order to present the linear program, we require a bit of notation.
For $q_1, \ldots, q_{k-1}, a_1, \ldots, a_k \in \set{0, \ldots, N-1}$
(the $q_{\ell}$'s representing a discrete sparc and the $a_{\ell}$'s
representing a cell), for a permutation $\sigma$ of $\set{1, \ldots,
  k}$, and for $i \in \set{1, \ldots, k}$, define 
\[
f_i^{\sigma}((q_1, \ldots, q_{k-1}), (a_1, \ldots, a_k)) = 
\left\{\begin{array}{ll}0 & \mbox{if }\exists m < \sigma^{-1}(i)\ :\
  q_m > a_{\sigma(m)}\\
1 & \mbox{otherwise}
\end{array}
\right.
\]
The value of $f_i^{\sigma}$ is 0 if some slice earlier than the slice
for terminal $i$ captures the cell defined by $a_1, \ldots, a_k$.

Define $\delta(m, n)$ to be 1 if $m=n$ and zero otherwise.

The linear program minimizes $\tau$ subject to the following
constraints.

\begin{equation}
\sum_{q_1, \ldots, q_{k-1}} x_{q_1, \ldots, q_{k-1}} = 1
\label{probability-distribution-equality}
\end{equation}

\begin{eqnarray}
\frac{1}{N^k} \sum_{q_1, \ldots, q_{k-1}} \frac{1}{k!} \sum_{\sigma} & &
(f_i^{\sigma}((q_1, \ldots, q_{k-1}), (a_1, \ldots, a_k))
\delta(q_{\sigma^{-1}(i)}, a_i) N \\
& + & 
f_j^{\sigma}((q_1, \ldots, q_{k-1}), (a_1, \ldots, a_k))
\delta(q_{\sigma^{-1}(j)}, a_j) N) x_{q_1, \ldots, q_{k-1}} \leq \tau
\end{eqnarray}

We can exploit symmetry to further reduce the number of constraints we
consider.  Since by assumption our sparc slices terminals in random
order, two segments that are identical under permutation of
coordinates will have the same densities, so we need consider only one
of them.  Thus, we restrict our constraints to $1,2$-aligned segments
in which the remaining coordinates are in nondecreasing order.  For
each such segment and cell, we use one constraint to measure the
average density induced by a given sparc over all permutations of the
terminal orders.

\subsection{LP Results}

Exploiting symmetry as discussed above, we were able to solve
relatively fine discretizations of the problem.  We wrote a simple
program to generate the linear programs automatically, and used CPLEX
to solve them.  While it is difficult to ``prove'' programs correct,
our computations did converge to the correct $12/11$ approximation
ratio for the 3-terminal case.

We give our results below in tabular form.  We derived improved bounds
for $4$--$9$ terminals.  Note that these programs optimize a proven upper bound on the approximation ratio; thus,under the assumption that the
programs were correct, these numbers are proven upper bounds.  In fact,
since the programs output a particular distribution over discrete
cuts, their performance ratio could be proven analytically via a
tedious case analysis on each cell of the discretize grid (which we
have not performed).

\ny{}{}{changed ``LP Gap'' to ``Bound'' in the column heading,
  cleaned up table.}
{\small
\begin{center}
\begin{tabular}{|r||c|c|c|c|}
\hline
  $k$ & \shortstack{grid\\size} & \shortstack{corner cut\\probability}&bound& $3/2-1/k$ \\
\hline
 3  &90   &.284 &1.0941   &  1.167  \\
 4  &36   & .289&1.1539   &  1.250     \\
 5  &18   & .314&1.2161   &  1.300      \\
 6  &12   & .376&1.2714   &  1.333  \\
 7  &9    & .397&1.3200   &  1.357    \\
 8  &6    & .414  &1.3322   &  1.375   \\
\hline
\end{tabular}
\end{center}
}
Our experiments also revealed one interesting fact: in all cases, the
optimum cut distribution made use of ``corner cuts.''   
That is, the output distribution had the following form:
with some probability, place each slice at a distance chosen uniformly
between 0 and $1/3$ from its terminal; otherwise, use a (joint)
distribution that places every slice at distance greater than $1/3$
from its terminal.

Adding constraints that forced the corner cuts to operate over a range other
than 1/3 of the way from the terminals worsened the computed
performance ratio, hinting that perhaps the optimal algorithm uses
corners of size exactly 1/3.
 This result is consistent with the optimal 3-terminal
algorithm, 
but inconsistent with the corner cut placement in the analytical
solution for higher $k$ that we give later.  We may be observing a
misleading artifact of working with a small discretized problem, or we
may be missing something in our analytic solution.

\section{Upper Bound for $k=3$}
\label{sec:ub3}

Our analytic upper bound of $12/11$ for $k=3$ comes from a new cutting
scheme that we call the ball/corner scheme.  Though for simplicity we
present a non-sparc scheme, there is a similar scheme using sparcs
that achieves the same bound.

For $k=3$, the simplex $\Delta$ can be viewed as a triangle in the
plane, which simplifies our pictures.  However, we continue to
use the original three-dimensional coordinate system to locate points
in the simplex.  Our cut of the simplex is determined by some lines
and rays drawn through the triangle; we refer to them as {\em
  boundaries}.  We will show that no segment has high density with
respect to our random choice of boundaries.

As illustrated in Figure~\ref{threeFig}, number the vertices of the
simplex $1,2$ and $3$.  Let points $a,b,\ldots,f$ divide the edges in
thirds, so that $a$--$b$--$f$--$d$--$c$--$e$--$a$ is the hexagon in
$\Delta$ with side length\footnote{Remember that we measure length as half
  the $L_1$ norm.}  $1/3$, with side $c$--$d$ on the side of the
triangle connecting terminals $2$ and $3$.  Note that this hexagon is
(a scaled version of) the unit ball for our distance metric.  The
points on the boundary of the hexagon are each at distance $1/3$ from
the hexagon's center.Outside the hex, we have a corner for each terminal
$i$ consisting of the points $x$ with $x_i>2/3$.

\input{threefig.tex}

\setlength{\unitlength}{0.35pt}
\begin{figure*}[tbp]
  \begin{center}
    \begin{tabular}{lcr}
      \begin{picture}(330,230)(0,0)
        \doTriangle
        \doTriangleLabels
        \doCorners
        %%\doTriangleLabels
        %%\doCorners
        \thicklines
        \doUnif
        \doUnifLabels
      \end{picture}  
      &
      \begin{picture}(330,230)(0,0)
        \doTriangle
        \doTriangleLabels
        \doCorners
        \thicklines
        \doBallCut
        \put(\PX,\PY){\makebox(0,0)[t]{\shortstack{~\\~\\$\Plabel$}}}
        %\doBallCutLabels
      \end{picture}  
      &
      \begin{picture}(330,230)(0,0)
        \doTriangle
        \doTriangleLabels
        %%\doTriangleLabels
        %%\doStarLabels
        \doCorners
        \thicklines
        \doCornerCut
        %\doCornerCutLabels
      \end{picture}
    \end{tabular}
  \end{center}
  \caption{This figure illustrates the cuts used for the case $k=3$.
    The ball is contained within the dotted lines.
The leftmost diagram shows how $r$ might be chosen for the ball cut.
The middle diagram shows one possible resulting ball cut (bold lines).  The
rightmost diagram shows a corner cut (bold lines).}
  \label{threeFig}
\end{figure*}
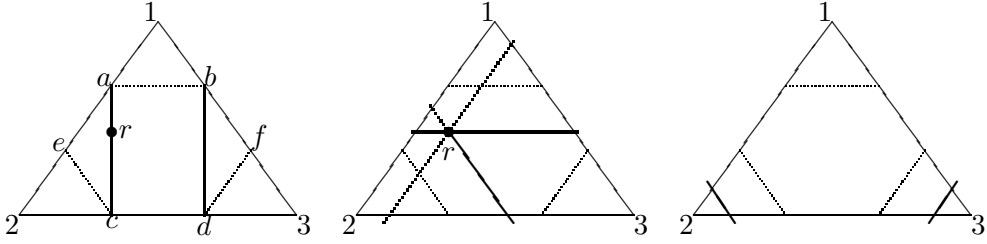

\subsection{The Ball/Corner Scheme}

The ball/corner scheme chooses a {\em ball cut\/} with probability $8/11$,
otherwise it chooses a {\em corner cut}.  These two types of cuts are
defined next.
The scheme is illustrated in Figure~\ref{threeFig}.

%%%KARGER Why are these undelined rather than normal boldface?
\noindent {\bf Ball cut:} Choose a point $r$ uniformly at random
from either line $a$--$c$ or line $b$--$d$.  
Consider the three
lines $\plane i {r_i}$ ($i=1,2,3$) parallel to the triangle's sides 
and passing through the point $r$.  Each such line is divided at the 
point $r$ into two rays.  Thus we get six rays.  Each
side of the triangle intersects two of these rays.  For each
side,  choose uniformly at random one of the two rays that hit it.  This
gives three rays; they form the boundary of the $3$-way cut.  (For a
sparc-based equivalent of this algorithm, we can choose (in random
order) two of the three sparcs that pass through the chosen point $r$.)

\noindent {\bf Corner Cut:} Choose two terminals in
$\set{1,2,3}$, and a value $\rho\in(2/3,1]$, uniformly at random.  For
each of the two chosen terminals $i$, let $l_i=\plane i {\rho}$.  The
two lines $l_i$ form the boundaries of the $3$-way cut.  (Note
  that a corner cut is a sparc.)

%% ney - don't think this really needs to be said,
%% because our analysis applies to any reasonable
%% interpretation of our description of the cutting scheme
\iffalse
Note that we do not specify where the points on the cutting lines
belong in the $3$-way cut of the simplex; these can be assigned to any
block since they have measure 0 relative to our densities.
\fi

\subsection{Analysis}  

We first state two simple properties of the ball cut that we need to
analyze the performance of the cutting scheme:
\begin{fact} Each of the $3$ coordinates of the random point $r$ is uniformly
  distributed in $[0,2/3]$.
  \label{fact:uniform}
\end{fact} 
\begin{fact} 
  Once $r$ is chosen, each one of the six candidate rays connecting
  $r$ to one side of the triangle is chosen with probability $1/2$.
  \label{fact:half}
\end{fact} 
\ny{}{\tt DK: Using an equilateral triangle instead of two lines for choosing
    $r$ makes this fact much easier to justify.  Can we
    change?}{Not sure equilateral triangle works. (I forget.)  We
  would have to redraw the pictures.}
\newmt{}{}{I am pretty sure that David is right about the equilateral
triangle. However, I am not sure it is easier to justify, and the
as Neal sais, we already have the other figures.}
  
\begin{theorem}
  The maximum density of the ball/corner scheme is $12/11$, so 
  $\tau_3^* \le 12/11$.
\label{lemma:3cut}
\end{theorem}
\begin{proof}
  We show that the expected density of any segment $e$ is at most
  $|e|\cdot 12/11$.  For the ball cuts, we use only the two facts
  claimed above.  Since these two facts, as well as the corner cut
  scheme, are symmetric with respect to the three coordinates, it
  suffices to prove the claim only for a $1,2$-aligned segment $e$.
  Further, we may assume assume that $e$ is entirely contained
  in either a corner or the hex; for otherwise, as discussed in
  Section~\ref{sec:density}, we can just split $e$ into corresponding
  pieces, calculating the density for each piece separately. We
  will consider several cases, depending on where $e$ is located.

  \iffalse Note that all boundary lines are contained in some $\plane
  i \alpha$.  This was already stated for the corner cuts.  For the
  ball cuts, the six possible lines out from $r$ are the two sides of
  each of the the three lines $\plane i {r_i}$ obtained by cutting at
  $r$. By Fact~\ref{basecut}, $e$ cannot be cut by $\plane i \alpha$
  if $i=3$.  \fi

  First, assume $e$ is located entirely in the hex.  Such a segment
  cannot be cut by a corner cut, so we need only consider the
  density when a ball cut is made and multiply by the probability of
  choosing a ball cut, namely $8/11$.  Assume a
  ball cut is made. Then $e$ can only be cut by rays
  in $\plane i {r_i}$ for $i=1,2$. By Fact~\ref{fact:uniform}, $r_i$
  is uniformly distributed in $[0,2/3]$.  Hence, the probability that
  $\plane i {r_i}$ goes through $e$ is
  $|e|/(2/3)$ since $e$ is $1,2$-aligned. If $\plane i {r_i}$
  touches $e$, it is at a single point. By Fact~\ref{fact:half}, 
  the ray of $\plane i {r_i}$ containing this point is picked for the
  cut with probability $1/2$.  Thus the expected number of times $e$
  is cut is
  $\frac{8}{11}\cdot2\cdot\frac{|e|}{2/3}\cdot\frac{1}{2} = \frac{12}{11}|e|$.

  Exactly the same argument applies if the edge is in the corner
  closest to terminal $3$.  The ball cut contributes the same $12/11$
  density, while the corner cut contributes nothing (note that a
  $1,2$-aligned edge is parallel to the line
  $\plane 3 {r_3}$, so cannot be cut by it).

  Finally, suppose segment $e$ is in the corner closest to terminal $1$
  (a symmetric argument applies if $e$ is in the corner closest to
  terminal $2$).  In this case, if a ball cut is made, the above
  %%ney 12/7/98
  \iffalse
  analysis applies except that only one of the two pieces of $\plane i
  {r_i}$ can cut $e$ (the other piece never enters a corner), so the
  contribution by a ball cut is halved to $|e|\frac{6}{11}$.  But the
  \fi
  analysis applies except that only the line $\plane 2 {r_2}$
  can cut $e$ (the line $\plane 1 {r_1}$  never enters the corner), so the
density  contribution of the ball cut is halved to $|e|\frac{6}{11}$.  But the
  %%end ney
  edge can also be cut by a corner cut.  A corner cut is chosen with
  probability $3/11$.  When it is, two of the three terminals are
  chosen, so terminal $1$ is chosen with probability $2/3$.  If
  terminal 1 is chosen, then, since the cutting line near terminal 1
  is of the form $\plane 1 {1-p}$, where $p$ is chosen uniformly in
  $[0,1/3]$, the probability that the line cuts $e$ is $|e|/(1/3)$.
  Thus, the expected number of times that the edge $e$ is cut (by a
  ball cut or corner cut) is
  $|e|\frac{6}{11}
  + \frac{3}{11}\cdot\frac{2}{3}\cdot\frac{|e|}{1/3} =
  |e|\frac{12}{11}$.

\end{proof}

\section{Lower Bound for $k=3$}
\label{sec:lb3}

\begin{theorem}
  For $k=3$, the minimum maximum density $\tau_3^* \ge 12/11$.
% CCC added this
Hence, the integrality gap for the geometric relaxation is $12/11$.     
\end{theorem}
Note that  this theorem applies to all cutting schemes, not just
sparcs.  Thus, the scheme of the previous section is optimal.

%%% CCC incorportated in theorem
\iffalse
\begin{corollary}
The integrality gap for the geometric relaxation is $12/11$.  
%The claim holds even when we restrict attention to planar graphs.
\end{corollary}
\fi 

\begin{proof}
  Fix $N$ to be any positive integer.  We construct an embedded
  weighted graph $G_N$ with no 3-way cut of cost less than $12N$, but
  with an embedding of cost $11N+1$.  This immediately
    demonstrates an integrality gap of $12N/(11N+1)$.  Furthermore,
    it implies that no cutting scheme has maximum density
  less than $12N/(11N+1)$, because by Lemma~\ref{densitylemma} such a
  cutting scheme applied to $G_N$ would yield a $3$-way cut with
  expected cost less than $12N$, a contradiction.  Since $N$ is
  arbitrary, the result follows.  Our construction (for $N=7$) is
  shown in Figure~\ref{lbfig}.

%%% CCC was 0.04 0.06 for scalebox, deleted shown
  \begin{figure*}[t]
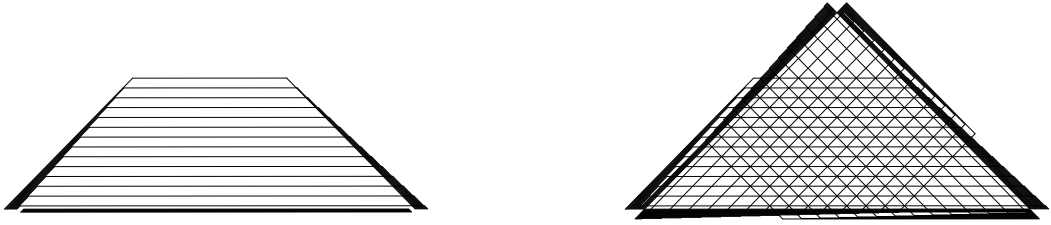

    \centerline{%
      \hfill
      \rotatebox{0}{\scalebox{0.03}[0.03]{\includegraphics{lb1.eps}}}
      \hfill
      \hfill
      \rotatebox{0}{\scalebox{0.03}[0.03]{\includegraphics{lb.eps}}}
      \hfill
      }
    \caption{The lower bound for $k=3$ (here $N=7$).
      The paths from $2$ to $3$ are on the left.
      The entire graph is on the right.
      On the border, overlapping paths are drawn side-by-side for clarity,
      so line width represents edge cost.}
    \label{lbfig}
  \end{figure*}

  For any pair of distinct terminals $i,j$ and number $d\in[0,1]$,
  define embedded path $p(i,j,d)$ as follows.
  Let $\ell$ be the terminal in $\set{1,2,3}-\set{i,j}$,
  let $a$ be the point on segment $i\ell$ at distance $d$ from $i$, and
  let $b$ be the point on segment $j\ell$ at distance $d$ from $j$.
  Then $p(i,j,d)$ is the union of the three segments $ia$, $ab$, and $bj$.
  
  We form the graph from $9N$ paths $p(i,j,d)$ for $0\le d \le 2/3$,
  where $d$ is an integer multiple of $1/(3N)$.  Although
  we describe the graph as a set of paths, technically it is a
  planar graph consisting of nodes and edges as follows: for every
  point in $\Delta$ whose coordinates are integer multiples of
  $1/(3N)$, there is a node in the graph embedded at that point; for
  every pair of nodes embedded $1/(3N)$ units apart, $G$ has an edge
  with cost equal to the number of paths that pass through both nodes.
  
  With this understanding, we now specify the graph.
  For each of the $3$ distinct pairs of terminals $i,j$, there are $3N$
  paths.  Of these paths, $N$ run directly between the terminals; that
  is, there are $N$ copies of $p(i,j,0)$.
  The remaining $2N$ paths are the paths $p(i,j,m/(3N))$
  where $m=1,2,\ldots,2N$.
  
  The total cost of the embedding is the total length of the paths.
  Since a path $p(i,j,m/(3N))$ has length $1+ m/(3N)$, a direct
%%%KARGER not clear why p() has specified length
  calculation shows that the total length of the paths is
  $3[N+\sum_{m=1}^{2N}1+m/(3N)] = 11N+1$.
  
  Next we lower bound the cost of any 3-way cut.  Since the graph is
  planar, any minimal 3-way cut corresponds either to a disconnected cut
  (meaning that the cut is the union of two disjoint 2-way cuts, each
  separating some terminal from both other terminals), like our upper
  bound's corner cut, or a connected cut
  (meaning that the cut edges give, in the planar dual, three paths
  connected at some central node and going to the three sides of the
  triangle), like our upper bound's ball cut.
  
  Any 3-way cut must cut all of the $9N$ paths at least once.  To
  finish the proof, we will argue that for either type of 3-way cut
  (connected or not), at least $3N$ paths are cut twice, so that
  the edges cut by the 3-way cut cost at least $12N$.
  This is easy to verify for a disconnected cut: a disconnected cut is
  the union of two 2-way cuts, so the $3N$ paths running between the
  two terminals that are cut off must be cut twice.
  
  Now consider any connected cut. In the planar dual of $G_N$, the
  connected cut corresponds to a central node and three paths from the
  node to each side of the triangle.  Let $x=(x_1,x_2,x_3)$ be any
  point inside the face of $G_N$ corresponding to the central node.
  Consider a path $p(i,j,d)$ such that $d \ge x_\ell$, where $\ell\neq
  i,j$.  That is, $x$ is inside the cycle formed by the union of
  $p(i,j,d)$ and $p(i,j,0)$.  Then the path $p(i,j,d)$ is cut at least twice by
  the connected cut.  For fixed $i$ and $j$, the number of such paths
  (with $d \ge x_\ell$) is $\lfloor 2N-x_\ell3N+1\rfloor > 2N-3x_\ell N$.
%%%KARGER why this many?
  Thus, the total number of such paths is more than
  $6N-3(x_1+x_2+x_3)N = 3N$.
\end{proof}

\section{Improvement for general $k$}
\label{sec:ub}

We now present an algorithm for an arbitrary number of terminals.
  While this algorithm seems unlikely to be the best possible, it
  improves on the previous best bound.  As discussed in
  Section~\ref{sec:sparc}, the essential observation in this analysis
  is that many slices can capture an edge before it has a chance to be
  cut. 

\newcommand{\ran}{\mbox{ICUT}}
\begin{theorem}\label{thm:gen-k} For all $k$, $\tau_k^* \leq 1.3438$.
Moreover, there is a $k$-way cut randomized approximation algorithm with an
approximation guarantee of $1.3438$.
 \end{theorem}
Our bound improves on the  Calinescu et al.
bound of $1.5-2/k$ for all $k\geq 14$.
For $k<14$, we show that $\tau_k^* < 1.5-2/k$
  by specializing the analysis for small $k$
  (see Subsection~\ref{sec:smallk}).

\newmt{%
To prove the theorem, we will use a (sparc) cutting scheme, that is,
we choose $k$ slicing thresholds $\rho_i$, and apply the slices
$\plane i {\rho_i}$ to a random permutation $\sigma$ of the
terminals. We are going to apply either an \emph{independent cut} ($\ran$) or
a corner cut: 
\\\noindent\underline{\ran:} each $\rho_i$ is chosen independently and
uniformly in $[0,6/11]$.
\\\noindent\underline{Corner cut:} all $\rho_i$ are chosen equal to a 
single random $\rho$ picked uniformly in $[6/11,1]$.
\\\noindent We will apply \ran\ with probability
$\alpha=0.667186$ and apply a corner cut with the remaining probability.

Before proving that the above sparc achieves a maximum density below
1.3438, as in Theorem \ref{thm:gen-k}, 
let's first draw parallels to our scheme for $k=3$. The corner
cut is completely analogous, except that $\rho$ is now chosen in the
interval $[6/11,1]$ instead of $[2/3,1]$. However, $\ran$ is very different
from the ball cut. For example, the $\rho_i$ are now independent whereas 
they were highly dependent in the ball cut. The reader may wonder why
we did not just generalize the ball/corner cut scheme. However, corner
cuts are only meaningful for $\rho=(1/2,1]$ since this is the region
in which the $\plane i {\rho}$ are disjoint. On the other hand,
the maximal ball of the $k$-simplex has center $(1/k,\ldots,1/k)$ and
radius $1/k$. Already for $k=4$, the simplex cannot be covered
by a ball and corners, and for $k$ large, the measure of the ball is
vanishingly small. Hence, the concept of ball cuts is not really relevant
for large $k$.
}{%
To prove the theorem, we will use a (sparc) cutting scheme
called \ran: we choose $k$ slicing thresholds $\rho_i$, and apply the
slices $\plane i {\rho_i}$ to a random permutation $\sigma$ of the
terminals (Note the difference from the 3-terminal case, in which
  the choices for the slice distances were dependent).  Each
  $\rho_i$ is chosen independently and uniformly in the interval
  $[0,6/11]$.%
}{I think it is important to start of telling the reader the full scheme
so that he can immediately appreciate that it is simple and easy to
implement, before dwelling into analysis. Also, I added the discussion
of why ball cuts don't work. I don't know when this discussion dissappeared,
but it must be the obvious question in any readers mind: ``why don't they just generalize.}

To bound the cutting density of our scheme, we will bound the 
density of every segment.  As justified in Section~\ref{sec:align},
we consider a segment of length $\epsilon>0$, and let $\epsilon$ approach zero.
As in the ball/corner scheme, by symmetry we can assume without loss
of generality that the segment is $1,2$-aligned.

Define $d_k(x_1,\ldots,x_k)$ to be the density with which \ran\ cuts
a $1,2$-aligned segment of infinitesimal length located at
$x_1,x_2,\ldots,x_k$.  We will show:
\begin{lemma}\label{ran-lemma}
  \[ {%\small
d_k(x_1,\ldots,x_k)
  \le \cases{2.014096 & if $x_1,x_2 \le 6/11$ \cr
    11/12 & otherwise.}
}
  \]
\end{lemma}
\newmt{}{%
The final cutting scheme chooses to \ran\ with probability
$\alpha=0.66719$ and 
otherwise chooses a corner cut.  The corner cut is chosen by 
the natural generalization of the scheme for $k=3$: choose
a value $\rho \in [6/11,1]$ (creating a density of 11/5 for each
corner edge).  The $k$-way cut consists of the hyperplanes 
$l_i=\plane i {\rho}$, for each $i$.  Note that the last corner cut
need not technically be made but it simplifies the analysis.}{said earlier}

By Lemma~\ref{ran-lemma}, this combined scheme gives a density of
  $2.014096\alpha$ for non-corner segments (since they are cut only if
  \ran{} is used, and then only with probability 2.014096), and a
  density of $(11/12)\alpha+(11/5)(1-\alpha)$ for corner segments
  (combining their probabilities of being cut by the two schemes), for
  a maximum density of $\max\{(2.014096)\alpha, (11/12)\, \alpha +
(11/5)(1-\alpha)\} \le 1.3438$, proving Theorem \ref{thm:gen-k}.  To
finish the proof of Theorem \ref{thm:gen-k}, it remains only to prove
Lemma~\ref{ran-lemma}.

ICUT's cumulative probability distribution function for any $\rho_i$
is $F(z) = \min\{(11/6)z,1\}$.  The corresponding probability density function is
$$
{%\small
 F'(z) = \left\{
  \begin{array}{ll}
    11/6 & \mbox{if } z \in [0,6/11]\\
    0 & \mbox{otherwise.}
  \end{array}
\right.
}
$$
Consider a $1,2$-aligned segment of length $\epsilon$ with one
endpoint fixed at
$x_1,x_2,\ldots,x_k$.  As $\epsilon$ goes to zero, the density of this segment
goes to
\begin{equation}
d_k(x_1,\ldots,x_k)
  \,= \, \frac{1}{k!} \sum_{\sigma}\bigg(
    F'(x_1) \prod_{i:\sigma(i) <  \sigma(1)} \left[1-F(x_i)\right]
    \, +\,\, F'(x_2) \prod_{i:\sigma(i) <
     \sigma(2)}\left[1-F(x_i)\right]\bigg)   \label{density-eqn}
\end{equation}
where the sum is over all $k!$ orderings of the terminals.  
This formula follows from Fact \ref{sparc-prop}. 
The first term measures the probability that the segment is cut
  by terminal 1, which happens if the slice for terminal 1 goes
  through the segment while all slices preceding terminal 1 in the
  ordering fail to capture the segment.  The second term similarly
  measures the probability that the segment is cut by terminal 2.
  Considering the slices for terminals other than 1 and 2 is the
  crucial element in improving the density bound of 3/2 for large $k$.
The formula assumes that $F$ is continuous around each $x_i$ and that
$F'$ is continuous in an open region around $x_1$ and $x_2$.  The
latter is not the case around $6/11$. However, as discussed Section
\ref{sec:density}, we may assume that all segments $e$ considered have
been subdivided so that for each $i=1,2$, either $\max x_i\leq
6/11$ or $\min x_i \geq 6/11$.
%%%KARGER technical detail: seems to be a problem since subdividing
%%%can never ensure x_i > 6/11, can only ensure x_i >= 6/11.

Note that $d_k(x_1,\ldots,x_{i},0,\ldots,0) = d_i(x_1,\ldots,x_i)$
(provided $i\ge 2$), because $x_j = 0$ implies terminal $j$ cannot
save the edge.  Note also that $d_k$ is symmetric with respect to
the variables $x_i$ for $i>2$.
Define
\begin{eqnarray*}
D_k(x_1, x_2) & \doteq & \max_{x_3,\ldots,x_k} d_k(x_1,x_2,\ldots,x_k) \\
C_k(x_1,x_2) & \doteq & d_k(x_1,x_2,c,\ldots,c)\\
&&\hspace{0.01in}\mbox{ where } c = (1-x_1-x_2)/(k-2), \\
D_\infty(x_1,x_2) &\doteq & \lim_{k\rightarrow\infty} D_k(x_1,x_2), \\
C_\infty(x_1,x_2) & \doteq & \lim_{k\rightarrow\infty}C_k(x_1,x_2).
\end{eqnarray*}
In these definitions, $(x_1,x_2,\ldots,x_k)$ is required
to lie in the $k$-simplex. 

$D_k(x_1,x_2)$ is the maximum density of any $1,2$-aligned infinitesimal segment
with an endpoint whose first two coordinates are $x_1,x_2$.
Note that the maximum is well-defined and achieved by some $x_3,\ldots,x_k$
because the simplex is closed under limits.
%%%KARGER simplex is closed, by we are relying on a claim about the
%%%``open ball'' of portion of simplex > 6/11, which is not closed
%%%under limits.

To understand \ran, our first goal is to characterize $D_k$.
We consider $C_k$ as it is one candidate for $D_k$.
\begin{lemma}
 $D_k(x_1,x_2) \leq D_{k+1}(x_1,x_2)$ for all $k$.
\end{lemma}

\begin{proof} 
% CCC
\iffalse
  For any $x_1,\ldots,x_k$,
  \[ {\small
d_k(x_1,  x_2, \ldots, x_k) = d_{k+1}(x_1,x_2,\ldots,x_k,0),}
\]
\fi
$\max d_k(x_1, \ldots, x_k) = \max d_{k+1}(x_1,\ldots,x_k,0) 
\le \max d_{k+1}(x_1,\ldots,x_k,x_{k+1})$.
\end{proof}

Thus for fixed $x_1,x_2$, $\langle D_2(x_1,x_2),D_3(x_1,x_2),\ldots \rangle$ 
is a nondecreasing sequence bounded from above (by 2). 
This implies that $D_\infty$ is well-defined.  We will see later that
$C_\infty$ is also well-defined.

Next we show that for fixed $x_1$ and $x_2$, the maximum for $d_k$ occurs at either
the ``central point'' $x_1,x_2,c,c,\ldots,c$ or the ``three-terminal'' point 
$x_1,x_2,1-x_1-x_2,0,\ldots,0$.
\begin{lemma} \label{same-lemma}
  \[
d_k(x_1,\ldots,x_k) \leq
    \cases{C_k(x_1,x_2) & if $\forall i>2: x_i\le 6/11$\cr
%%ney 2/6/99, replaced > with >=
%   C_3(x_1,x_2)& if  $\exists i>2:x_i>6/11$}$
      C_3(x_1,x_2)& if  $\exists i>2:x_i\ge6/11$.}
\]
\end{lemma}

\begin{proof} 
  Fix $x_1$ and $x_2$.  Let $c=(1-x_1-x_2)/(k-2)$.
  
  Claim 1: {\em Among all $x_3,\ldots,x_k$ such that $0 \le x_i \le 6/11$ for
    all $i>2$ (and $x_1,x_2,\ldots,x_k$ is in the simplex), the unique
    maximizer of $d_k(x_1,x_2,x_3,\ldots,x_k)$ satisfies
    $x_3=x_4=\cdots=x_k$, so is equal to $C_k(x_1,x_2)$.}  Suppose for contradiction that some other such $x_3,x_4,\ldots,x_k$ maximizes
  $d_k$.  Then $x_i < x_j$ for some $i,j > 2$.  Considered just as a function
  of $x_i$ and $x_j$ (holding the other coordinates fixed)
%  {\small 
\begin{equation}
d_k(x_1,\ldots,x_k)
\,=\, p \, +\, q [1-F(x_i)] + r [1-F(x_j)]\,  + \, s    [1-F(x_i)][1-F(x_j)] 
 \label{density-i,j}
\end{equation}
\iffalse
\begin{eqnarray}
\nonumber    {d_k(x_1,\ldots,x_k)}
%\\
\nonumber & = & p + q [1-F(x_i)] + r [1-F(x_j)] \\
\mbox{} &&+ s    [1-F(x_i)][1-F(x_j)] 
 \label{density-i,j}
  \end{eqnarray}
\fi
%}
  where $p$, $q$, $r$ and $s$ are nonnegative and independent of $x_i$ and
  $x_j$.  Furthermore $q=r$ because $d_k$ is symmetric in $x_i$ and $x_j$.
  Consider increasing $x_i$ and decreasing $x_j$ at equal rates.  This
  maintains $0 \le x_i,x_j \le 6/11$ but increases $d_k$ at a rate
proportional to
\[
q [F'(x_j)-F'(x_i)]\, + \, s \Big(F'(x_j) [1-F(x_i)] - F'(x_i)[1-F(x_j)]\Big).
%\label{rate-of-change}
\]
\iffalse
{\small
  \begin{eqnarray*} 
\nonumber    \lefteqn{q [F'(x_j)-F'(x_i)]}\\
& \mbox{} + s \Big(F'(x_j) [1-F(x_i)] - F'(x_i)[1-F(x_j)]\Big).
%\label{rate-of-change}
  \end{eqnarray*}
  }
\fi
  This is positive because $F'(z) = 11/6$ for $z< 6/11$ and $F(x_j) > F(x_i)$
  (recall that $x_i<x_j\le6/11$).  This contradicts the choice of
  $x_3,\ldots,x_k$.
  
  Claim 2: {\em Among all $x_3,\ldots,x_k$ such that $x_i \ge 6/11$ for some
    $i>2$ (and $x_1,\ldots,x_k$ is in the simplex), the unique maximizer of
    $d_k(x_1,x_2,x_3,\ldots,x_k)$ satisfies $x_i = 1-x_1-x_2$ and $x_j = 0$ for $j\neq
    i$, so is equal to $C_3(x_1,x_2)$.}   Suppose for contradiction that some other such
  $x_3,x_4,\ldots,x_k$ maximizes $d_k$.  Fix some $j>2$ such that $0<x_j<6/11\le
  x_i$.  Since by assumption $x_i ge 6/11$, we have $F(x_i)=1$ and thus the expression~(\ref{density-i,j}) reduces to
  $p+r(1-F(x_j))$.  If we increase $x_i$ and decrease $x_j$ at the same rate, the
  rate of increase in $d_k$ is $r F'(x_j) > 0$, contradicting the choice
  of $x_3,\ldots,x_4$.

  The two claims together prove the lemma.
\end{proof}

\begin{lemma}
%%%ney 2/6/99 ``<'' is false for x_1+x_2=1
%  For $k \ge 4$, $C_k(x_1,x_2) < C_{k+1}(x_1,x_2)$.
  For $k \ge 4$, $C_k(x_1,x_2) \le C_{k+1}(x_1,x_2)$.
\end{lemma}

%ney
\begin{proof}
\vspace*{-.3in} %%% CCC hack
\begin{eqnarray*}
  \ C_k(x_1,x_2)
  &= &d_k(x_1,x_2,c,\ldots,c)\\
  &= &d_{k+1}(x_1,x_2,c,\ldots,c,0)\\
  &\le &C_{k+1}(x_1,x_2).
\end{eqnarray*}
Here $c=(1-x_1-x_2)/(k-2)$.
The last inequality follows from
Lemma~\ref{same-lemma} (using $c\le 1/2<6/11$).
\iffalse
  But by Claim 1 in the proof of Lemma~\ref{same-lemma} (using the fact that $c
  < 6/11$),
  $$d_{k+1}(x_1,x_2,c,\ldots,c,0) <
  d_{k+1}(x_1,x_2,c',\ldots,c',c') = C_{k+1}(x_1,x_2).$$
\fi
\end{proof}

An immediate corollary is that
$C_{\infty}(x_1,x_2)$ is well-defined and
$C_k(x_1,x_2) \le C_{\infty}(x_1,x_2)$ for all $k$.
Using this and Lemma~\ref{same-lemma},
to bound $D_\infty$ it suffices to bound $C_3$ and $C_\infty$.  
We begin with $C_\infty$.

\begin{lemma}\label{C-lemma}
%%%ney 2/6/99 - why was this line in there?
%%%  If $c=(1-x_1-x_2)/(k-2)\leq 11/6$,
  $$
{
C_\infty(x_1,x_2)
  \le \cases{2.014096 & if $x_1,x_2 \le 6/11$ \cr
    11/12 & otherwise.}
}
  $$
\end{lemma}

\begin{proof}
  %%ney - 
  \iffalse
  First consider the case $x_1+x_2 = 1$.  By inspection
  $C_k(x_1,x_2) = C_2(x_1,x_2)$ which is
  $$\frac{1}{2}\,F'(x_1)[2-F(x_2)]
  +\frac{1}{2}\,F'(x_2)[2-F(x_1)]
  $$
  If $x_1,x_2 \le 6/11$ then
  this simplifies to $(1/2)(11/6)[4-(11/6)] < 2$.
  Otherwise (e.g.\ when $x_1 > 6/11$,
  $F'(x_1) = 0$ and $F(x_1) = 1$)
  it is $(1/2)(11/6) = 11/12$.
  \fi
  Fix $x_1$ and $x_2$.
  Our first goal is to derive a closed-form
  expression for $C_k(x_1,x_2)$ for any $k$.
  Fix $k$ for now and let $x_i = c = (1-x_1-x_2)/(k-2)$ for $i>2$.
  
  For $j=1,2$, let $S_j$ denote the probability that the segment at
  $(x_1,x_2,\ldots x_k)$
  is not captured by a terminal other than $j$ before terminal $j$'s cut is made:
  $${
S_j \doteq \frac{1}{k!}\sum_{\sigma} \prod_{i:\sigma(i)<\sigma(j)} \left(1-F(x_i)\right).}$$
  Then $C_k(x_1,x_2) = S_1 F'(x_1) + S_2 F'(x_2)$.

  We will derive a closed-form expression for $S_1$
  (and by symmetry for $S_2$).
  Recall that $x_i=c$ for $i>2$.  We thus rewrite
%%% CCC commented out  Our first observation is
\[
    S_1 \,=\, \frac{1}{k}\sum _{q=0}^{k-1}\,
\left(\frac{q}{k-1} \,[1-F(c)]^{q-1}[1-F(x_2)] \,+\,\Big(1-\frac{q}{k-1}\Big)\, [1-F(c)]^q\right).
\]
  Here we condition on $q$, the number of $i$
  such that $\sigma(i) < \sigma(1)$.  Note that $q$ is uniform in $\{0,1,\ldots,k-1\}$
  while $\frac{q}{k-1}$ is the probability that $\sigma(2)<\sigma(1)$,
  given $q$. 
  A change of variables and rewriting give
  $$
  S_1 \, =\,
  \left(1+\frac{1-F(x_2)}{k-1}\right)\,\sum_{q=0}^{k-2}\frac{[1-F(c)]^q}{k}
  \, -\,     F(x_2)\,\sum_{q=0}^{k-2}\frac{q[1-F(c)]^q}{k^2-k}.
  $$
  Now we let $k\rightarrow \infty$.  The two sums above have standard closed
  forms that tend respectively to
  $${
[1-e^{-a}] a^{-1}~~\mbox{and}~~[1-(1+a)e^{-a}]\,a^{-2},}$$
  where $a \doteq \lim_{k \rightarrow\infty}k\,F(c)=(1-x_1-x_2)F'(0)$.
  Thus, as  $k\rightarrow \infty$,
  $${ S_1 \rightarrow [1-e^{-a}] a^{-1}
  - F(x_2)[1-(1+a)e^{-a}]\,a^{-2}.}$$
  Of course $S_2$ tends to the above with $x_1$ replacing $x_2$.
  This gives us our closed-form expression for $C_\infty(x_1,x_2)$:
  \begin{equation}
    C_\infty(x_1,x_2) \,=\,
    [F'(x_1)+F'(x_2)]\times\frac{1-e^{-a}}{a}
    \,-\, [F'(x_1) F(x_2) + F'(x_2) F(x_1)]
    \times\frac{1-(1+a)e^{-a}}{a^2}.
    \label{eq:Cinf}
  \end{equation}
  where $a = (1-x_1-x_2)F'(0)$.
    
\iffalse
  Thus, $C_k(x_1,x_2)$ is
  \begin{eqnarray*}
    \frac{1}{k}\sum _{q=0}^{k-1}
    \left[\frac{q}{k-1}(1-F(x_2)) (1-a/k)^{q-1} \right.\\
    \quad\left. \mbox{}     +\frac{k-1-q}{k-1} (1-a/k)^q \right] F'(x_1)+  \\
    \left[\frac{q}{k-1}(1-F(x_1)){{(1-a/k)}^{q-1}}\right.\\
    \left.  \quad\mbox{}    +\frac{k-1-q}{k-1}{{(1-a/k)}^q}\right]F'(x_2).
  \end{eqnarray*}
  In~(\ref{density-eqn}), $d_k$ is expressed as a sum over all permutations $\sigma$.
  In the expression above, $q$ represents the number of $j$ such
  $\sigma(j)<\sigma(1)$; $q/(k-1)$ is the chance that $\sigma(2)<\sigma(1)$.
\begin{eqnarray*}
\lefteqn{
\frac{n}{a^2 (n-1)} ((a-1+(1-\frac{a}{n})^n)
(F'(x_1)+F'(x_2))}\\
\lefteqn{-\frac{1}{1-a/n} (-1 + (a+1)(1-\frac{a}{n})^n +
\frac{a}{n}(1 - (1-\frac{a}{n})^n))}\\
&&\cdot (F'(x_1)+F'(x_2)-F(x_2)F'(x_1)-F(x_1) F'(x_2)))
\end{eqnarray*}

  \begin{eqnarray}
    %%C_\infty(x_1,x_2) =
\nonumber \lefteqn{    \frac{1}{a} (1-e^{-a})\big[F'(x_1)+F'(x_2)\big]}\\
\nonumber &&    - \frac{1}{a^2}(1-(1+a)e^{-a})\big[F'(x_1) F(x_2)\\
&&\qquad\qquad\qquad + F'(x_2) F(x_1)\big].
\label{eq:Cinf}
  \end{eqnarray}
\fi
The above equality holds for any suitably well-behaved $F$.
Using this closed form and our particular choice of $F$,
we now show the two desired bounds on $C_\infty$.
\paragraph{Case 1:}  $x_1,x_2\le 6/11$.
In this case $a=11/6(1-x_1-x_2)$, $F'(x_1)=F'(x_2)=11/6$,
and $F(x_1)+F(x_2)=11/6(x_1+x_2)=11/6-a$.
So (\ref{eq:Cinf}) gives
\begin{eqnarray*}
  C_\infty(x_1,x_2) \, = \,
  11/3\,{\frac {1-{e^{-a}}}{a}}
  \,-\,{\frac{121}{36}}\Big(1-{\frac {6}{11}}\,a\Big )
  \frac{1-\left (1+a \right ){e^{-a}}}{a^2}
\end{eqnarray*}
where $a=11/6\,(1-x_1-x_2)$ so $a\in[0,11/6]$.
Let $C(a)=C_\infty(x_1,x_2)$.  
In the rest of this case (Case 1),
we will prove that $C(a)\leq 2.014096$ for $a\in(0,11/6)$.
The cases $a=0$ and $a=11/6$ follow by the continuity of $C$.
The claim is ``obvious'' from a plot but the somewhat technical proof
appears below.

\iffalse
It suffices to show that $C'$ has a unique zero $a_0$, $C'(a)>0$ for $0<a<a_0$ and 
$C'(a)<0$ for $a>a_0$, and $C'$ is decreasing for $a>a_0$. 
\fi

We show that $C(a)$ is strictly concave for $a \in (0,11/6)$.  It
therefore has a unique maximum at some $a_0$, where $C'(a_0)=0$.
By substitution, $C'(.294) \ge 0.00045>0$ and $C'(.295)\le-0.00009<0$, so 
$a_0\in(.294,.295)$.  Hence 
\[\max_{a\in[0,11/6]}C(a)\leq C(.295)-0.001 \cdot C'(.295) \le 2.014096\]
To show $C(a)$ is strictly concave, we show that $C''(a)$ is strictly
negative. Now, 
%%% CCC replaced by below 
\[C'(a)\,=\,{\frac {11}{36}}\,\frac
  {7\,{e^{-a}}{a}^{2}-18\,a-4\,{e^{-a}}a}{a^3}
+\frac{11}{36}\,\frac{6\,{e^
    {-a}}{a}^{3}+22-22\,{e^{-a}}}{{a}^{3}}
\]
and
\[
C''(a)\,=\,-{\frac {11}{36a^4}}\,({7\,{e^{-a}}{a}^{3}}+3\,{e^{-a}}{a}^{2}-36\,a
-30\,{e^{-a}}a+6\,{e^{-a}}{a}^{4}+66-66\,{e^{-a}}).
\]
To show that $C''(a)$ is negative, it suffices to prove that
\[
D(a)\,=\,-7\,{e^{-a}}{a}^{3}-3\,{e^{-a}}{a}^{2}+36\,a+30\,{e^{-a}}a
-6\,{e^{-a}}{a}^{4}-66+66\,{e^{-a}}
\]
 is negative. By substitution,
$D(0)=0$ and $D(11/6)=0$, so it suffices to
show that  
$D'$ has only one zero $a_1$, $D'(a)<0$ for $a<a_1$ and
$D'(a)>0$ for $a>a_1$. Here
\[
D'(a)\,=\,-17\,{e^{-a}}{a}^{3}-18\,{e^{-a}}{a}^{2}-36\,{e^{-a}}a+36
-36\,{e^{-a}} +6\,{e^{-a}}{a}^{4}
\]
and
%%% CCC replaced by below
\iffalse
{\small
\[D''(a)=e^{-a}a^2 (-6a^2+41a-33)\]
}
\fi
$D''(a)=e^{-a}a^2 (-6a^2+41a-33)$.
For $a\in(0,11/6]$, $D''$ has only one zero
$a_2=\frac{41-\sqrt{889}}{12}\approx 0.93$
and $D''(a)<0$ for $a<a_2$ and $D''(a)>0$ for $a>a_2$. That is, $D'$ is
first decreasing and then increasing. Since $D'(0)=0$ and 
$D'(11/6)\ge 4.108>0$ it follows that $D'$ has only one zero $a_1$ 
for $a\in(0,11/6]$.

\paragraph{Case 2:}  $x_1$ or $x_2\ge 6/11$.
Assume $x_1 \ge 6/11$ (the case $x_2 \ge 6/11$ is
symmetric).  In this case, $F'(x_1) = 0$ and $F(x_1) = 1$,
  so we get
  $$C_\infty(x_1,x_2) =
{\frac {11}{6}}\,{\frac {1-{e^{-a}}}{a}}-{\frac {11}{6}}\,{\frac {1-
\left (1+a\right ){e^{-a}}}{{a}^{2}}}
$$
As before, let $C(a)= C_\infty(x_1,x_2)$.  We will prove that
$C(a)\leq 11/12$ for $a\in[0,11/6]$.   
First, $\lim_{a\rightarrow 0}C(a)=11/12$, so
$C(a)\leq 11/12$ follows if we can show that $C'(a) \leq 0$ for 
$a\in (0,11/6]$.  We have
\[C'(a) = \frac{11}{6a^3}(-a-{e^{-a}}a+2-2\,{e^{-a}}).\]
Define $E(a)=-a-{e^{-a}}a+2-2\,{e^{-a}}$. Since 
$\frac{11}{6a^3}>0$ for $a>0$, $C(a)\leq 0$ if and only if 
$E(a)\leq 0$.  
Since $E(0)=0$, we can infer $E(a) \leq 0$ if $E'(a) \leq 0$ for all $a
\in (0,11/6]$.  
We have 
%%% CCC replaced by below
\iffalse
\[{\small E'(a)=-1+e^{-a}(a+1).}\]
\fi
$E'(a)=-1+e^{-a}(a+1)$.
Note that $E'(0)=0$, so $E'(a)\leq 0$ follows if $E''(a) \leq 0$ for
$a \in (0,11/6]$.  We have
%%% CCC replaced by below
\iffalse
\[{\small E''(a)=-e^{-a}a}\]
\fi
$E''(a)=-e^{-a}a$,
so $E''(a)\leq 0$.   We conclude that 
$C_\infty(x_1,x_2)\leq 11/6$ if $x_1>6/11$.
\end{proof}

Lemmas~\ref{same-lemma} through~\ref{C-lemma} prove that,
for $x$ such that $x_i \le 6/11$ for all $i>2$,
{\small
\begin{eqnarray*}
d_k(x_1,\ldots,x_k)
& \le &C_\infty(x_1,x_2) \\
&\le &\cases{2.012096 & if $x_1,x_2 \le 6/11$ \cr
    11/12 & otherwise.}
\end{eqnarray*}}
The remaining case is when $x_i \ge 6/11$ for some $i>2$.
In this case by Lemma~\ref{same-lemma},
$${ d_k(x_1,\ldots,x_k)\le C_3(x_1,x_2) = d_3(x_1,x_2,1-x_1-x_2)}$$
and $x_1+x_2 \le 5/11$.
Thus, to finish the proof of the theorem, it suffices to show the
following lemma.
\begin{quote}
%{\bf Am I not doing the Lemma below Right???}
\end{quote}
\begin{lemma}\label{C3'-lemma}
 If $x_1+x_2 \le 5/11$, then  $C_3(x_1,x_2) \le 11/6 \le 2.012096.$
\end{lemma}
\begin{proof}
  Let $x_3 = 1-x_1-x_2 \ge 6/11$.

  Then $F(x_3) = 1$
  while $F(x_1) = 11/6\, x_1$,
  $F(x_2) = 11/6 \, x_2$,
  and  $F'(x_1) = F'(x_2) = 11/6$.

  By inspection of (\ref{density-eqn}),
  $C_3(x_1,x_2) = d_3(x_1,x_2,x_3)
  = (1/6)\,(11/6)\,(6-11/6\,(x_1+x_2)) \le 11/6$.
\end{proof}

This proves Lemma~\ref{ran-lemma}.

\subsection{Improvements for small values of $k$}
\label{sec:smallk}

For particular values of $k$ it is possible to refine the analysis in the proof
of Theorem~\ref{thm:gen-k} to get improved bounds.  In this case it is useful
to modify the algorithm so that it only uses $k-1$ cuts instead of
$k$.  In particular, we do not use the cut for the terminal $j$
with $\sigma(j)=k$.  The analysis for this modified algorithm goes similarly, with
our definitions appropriately modified to reflect that we are using
$k-1$ instead of $k$ cuts.

\iffalse
Then, all of the preliminary lemmas up to
Lemma~\ref{C-lemma} carry over, and for any corner size (not just 6/11) greater
than 1/2.  The first part of the analysis of Lemma~\ref{C-lemma} is easy to
adapt to the modified algorithm, and 
\fi
Then, instead of passing to the limit,
$C_k(x_1,x_2)$ can be evaluated directly.  Following this approach we 
obtained the following performance guarantees for particular $k$:
\begin{center}
\begin{tabular}{|r||c|c|c|} \hline
  $k$ & \shortstack{corner\\placement} & \shortstack{\ran\\probability} & bound \\ \hline   
  3 &.641& .675 & 1.131\\   
  4 &.607& .663 & 1.189\\ 
  5 &.588& .659 & 1.223\\ 
  6 &.576&.659 & 1.244 \\ 
  7 &.565&.657 & 1.258\\  
  8 &.557& .656 & 1.269\\ 
  9 &.557& .659 & 1.277  \\ 
  10 &.557& .661 & 1.284\\  
  12 &.554& .661 & 1.293\\  
  20 & .554 & .666 & 1.314 \\
  35 & .550 & .666 & 1.327 \\ \hline
\end{tabular}
\end{center}
``Corner placement'' is the placement of the corner (analogous
to $6/11$) and ``\ran{} probability'' is the probability of choosing \ran.
These parameters were chosen to try to minimize the resulting
bound on the performance ratio, shown under ``bound''.
These numbers are approximate;
the ratios were evaluated numerically without formal verification.

%\end{tabular}

\section{Conclusion}

We have provided a better analysis of an embedding relaxation for
multiway cut.  We have exactly determined the integrality gap for the
3-terminal problem, and given an approximation algorithm achieving the
bound.  For larger values of $k$, we have defined a class of cutting
schemes called sparcs that, through a combination of nonuniform and
dependent rounding, provide better approximation ratios than the best
previous schemes.  However, the question of the exact integrality gap
remains open.

%%% CCC commented out discussion
\iffalse
\section{Discussion}
This is ongoing research. We can obtain better bounds using more messy
approaches, but we do not present them here.
\fi
\iffalse
We already know we can do better using more
messy approaches. Perhaps we should just discuss 
Karger's experiment here instead of in a full section. 
\fi

%%% CCC

\section{Appendix}

\newcommand{\gap}{{\tt gap}}
\newcommand{\perf}{{\tt perf}}
\newcommand{\s}{{S}}
\newcommand{\FONT}{}
\newcommand{\G}{{\FONT G}}
\newcommand{\E}{{\cal E}}
\newcommand{\V}{{\FONT V}}
\newcommand{\W}{{\FONT W}}
\renewcommand{\W}{{\FONT w}}
\newcommand{\X}{{\FONT x}}
\newcommand{\Y}{{\FONT y}}
\renewcommand{\r}{{R}}
\newcommand{\IP}{{\cal IP}}
\newcommand{\LP}{{\cal LP}}

{\em Proof of Theorem \ref{infinite_case}}. 

\mtcom{Great job Neal. However, your use of $p/P$ and of $S$ is
not quite consistent. In the preceeding part of the paper, we always
use $P$ for cutting schemes, but below you use $p$ in the beginning and
$P$ later. Unless there is a good reason for this, I suggest using
$P$ everywhere. Also, concerning $S$, it mostly means solution, but in Lemma
7.3, you start using it on subsequences.}

We show that there necessarily exists a
cutting scheme whose maximum density equals the integrality
gap of the relaxation.  The basis of this proof is that the
two quantities are solutions to dual linear programs.
Although the linear programs are infinite, we show they have
no duality gap.

Interestingly, most of the proof holds in the following more
general setting: We have a non-negative real vector $x$ representing
a relaxed solution to some problem.   There is a set $\s$ of allowable
solutions (also nonnegative real vectors), and we want to round
$x$ to some solution $y \in \s$.  The method for rounding $x$ is
represented by a {\em randomized rounding scheme}, which is simply a
probability distribution $P$ on $\s$.

We assume that the {\em cost} of $x$ is given by $w\cdot x =
\sum_i w_i x_i$ for some nonnegative weight function $w$, and likewise the
cost of any $y\in\s$ is $w\cdot y$.  We want to choose a {\em
  single} rounding scheme $P$ that has a good performance
ratio against {\em all} possible weight functions $w$.
(We will see that this is analogous to choosing a {\em single}
rounding scheme of the simplex that has good performance
ratio against all embedded graphs.)

\mtcom{Added $\in p$ in the subscript of $E$,}
Define the performance ratio of a rounding scheme $P$ to be
\[\sup_w \frac{E_{y\in P}[w\cdot y]}{w\cdot x}
.\]
(This corresponds to the maximum density of a cutting scheme.)
Here the notation $E_{y\in P}[w\cdot y]$ signifies the expectation
over $y\in S$ chosen according to the probability distribution $P$.

Define the {\em integrality gap} to be
\[\sup_w \frac{\inf_{y\in \s} \,w\cdot y}{w\cdot x},\]
the worst-case ratio of the minimum-cost of any true
solution to the cost of the relaxed solution.

\begin{lemma}The performance ratio can be reformulated as
\[\sup_w \frac{E_{y\in P}[w\cdot y]}{w\cdot x}
\,=\, \sup_i \frac{E_{y\in P}[y_i]}{x_i}.\]
\end{lemma}

In the case of k-cut, each index $i$ corresponds to a
``seglet'' (edge of the divided simplex), and $w$ corresponds
to an embedded graph ($w_i$ corresponds to the number, or
total weight, of edges embedded along seglet $i$).  In that
case the lemma says that, to check the performance guarantee
of a rounding scheme, it suffices to check it for each seglet.

The proof is similar to the observation regarding two-player
zero-sum matrix games that once one player has fixed their
mixed strategy, the other player has an optimal mixed strategy
that is pure.

\medskip

\begin{proof}
  Clearly the left-hand side is greater than or equal to the
  right-hand side (take $w$ to be any of the $i$ unit vectors
  with a single coordinate equal to 1 and the others 0).  To
  finish we show that the left-hand side is at most the right
  hand side for any fixed $w$.

  Fix $w$.
  Let $\lambda$ equal the right-hand side above.
  Note that by linearity of expectation, the left-hand side is
  \[
  \frac{\sum_i w_i E_{y\in P}[y_i]}{w\cdot x}.
  \]
  By the definition of $\lambda$,
  $E_{y\in P}[y_i] \le \lambda x_i$, so the quantity above is at most
  \[
  \frac{\sum_i w_i \lambda x_i}{w\cdot x}
  \,=\, \lambda.
  \]
\end{proof}

\begin{theorem}\label{finite_case}
  If $\s$ and the dimension of $x$ are finite, then there
  exists a rounding scheme $P$ whose performance ratio
  equals the integrality gap.
\end{theorem}
\begin{proof}
Let $P_y$ be the probability that we choose solution $y$.
Choosing an optimum rounding scheme
is equivalent to the following linear program: 
\begin{linearprogram}{minimize$_P$}{\tau}
   \sum_{y\in \s} P_y \frac{y_i}{x_i} &\le& \tau & (\forall i) \\
   \sum_{y\in \s} P_y &=& 1 & \\
   P_y &\ge& 0 & (\forall y\in\s).
\end{linearprogram}
The dual of this program is
\begin{linearprogram}{maximize$_q$}{\lambda}
   \sum_i q_i \frac{y_i}{x_i} &\ge& \lambda & (\forall y\in\s) \\
   \sum_i q_i &=& 1 & \\
   q_i &\ge& 0 & (\forall i).
\end{linearprogram}
By the change of variables $w_i = q_i / x_i$, this
is equivalent to
\begin{linearprogram}{maximize$_w$}{\lambda}
   \sum_i w_i y_i &\ge& \lambda & (\forall y\in\s) \\
   \sum_i w_i x_i &=& 1 & \\
   w_i &\ge& 0 & (\forall y\in\s).
\end{linearprogram}
But it is easy to verify that this is equivalent to
the problem of choosing a weight function $w$
to achieve the integrality gap.
If $x$ and the vectors in $\s$ are finite-dimensional
and $\s$ is finite, then strong duality implies that
the linear programs have equal values.
\end{proof}

\newcommand{\EE}{\overline{\E}}
Next we describe how this relates to the rest of the paper.
Define edge set $\EE$ to be all edges (pairs of points) in the
simplex.
Define $x_e = |e|$ for $e\in\EE$.
Define $\s$ to contain the characteristic vectors $y(C)$ of
$k$-way cuts $C$ of the simplex:  $y_e(C) = 1$ if
$e$ is cut by $C$ and 0 otherwise. 
An embedding of a particular weighted graph $G$ in the simplex
then corresponds to a particular weight function $w(G)$, where
$w_e(G)$ equals the total weight of edges embedded on simplex edge $e$.
With this $x$, $\s$, and interpretation of $w$, we have the
following correspondences:
\begin{center}
\begin{tabular}{|rcl|} \hline
  general setting &  & $k$-way cut \\ \hline
  $w$ & $\leftrightarrow$ & embedded $G$\\
  $w \cdot x$ & $=$ & $\vol(G)$\\
  $y\in\s$ & $\leftrightarrow$ & $k$-way cut of $G$\\
  $w \cdot y$ & $=$ &
  cost of $k$-way cut $C$ of $G$\\
  rounding scheme $P$ & $\leftrightarrow$ & cutting scheme $P$\\
  $\inf_P$ performance ratio of $P$ & $=$ &
  $\tau^*_k$ \\
  integrality gap & $=$ &
  integrality gap \\ \hline
\end{tabular}
\end{center}

Thus, if $\s$ and the dimension of $x$ were finite,
we could conclude by the theorem that the
$\tau^*_k$ equals the integrality gap of the relaxation.
\dk{This yields an immediate corollary:

\begin{corollary}
For input graph instances of any bounded size, there is a rounding
scheme whose performance is equal to the integrality gap of the
relaxation.
\end{corollary}
\begin{proof}
  Consider the set of all graphs whose size is bounded by some
  quantity.  Each has a bounded number of vertices, and the (rational)
  weights on the graph are also of bounded size.  The linear
  programming relaxation thus also has bounded size.  It follows that
  any vertex solution to the linear programming relaxation, which assigns
  embedding coordinates to all the vertices, has bounded
  size---meaning that the coordinates are rational numbers of bounded
  size.  The set of coordinates at which vertices might be located in
  an optimal embedding therefore forms a discrete grid within the
  simplex, with a finite number of points.  The embedded edges connect
  these points, so there is a finite number of embedded edges in the
  optimal solution.  An input instance is determined entirely by the
  weights assigned to these edges, so has finite dimension.
  Similarly, for the purposes of rounding we need only consider
  $k$-way partitions of the finitely many grid vertices.  There are
  only finitely many of these partitions.

  Since the dimension of inputs and the number of output solutions is
  finite, the previous theorem applies and shows that there is a
  rounding scheme with performance equal to the integrality gap  of
  the embedding.
\end{proof}

If we want a rounding scheme that works (uniformly) for graphs of
arbitrary size, we have to work somewhat harder.  We}%
{Unfortunately, they are not finite. However, we}
we show that the desired result follows as a limiting case\newmt{}{.}{}
of Theorem~\ref{finite_case}.

In what follows, we restrict the $k$-way cut problem to various
particular subsets $E$ of the edge set $\EE$ of the simplex.
A $k$-way cut of $E$ is defined to be a
subset $C$ of $E$ such that $E-C$ contains no
terminal-to-terminal path
By the minimal maximum density {\em with respect to $E$}
we mean
\[\tau(E) = \inf_P \sup_{e\in E} \Pr[P \mathrm{\,cuts\,} e]/|e|,\]
where $P$ ranges over cutting schemes of $E$ (probability
distributions over $k$-way cuts of $E$).  By the integrality gap {\em with
  respect to $E$} we mean
\[\gap(E) = \sup_w \inf_C
\frac{\sum_{e\in C} w(e)}{\sum_{e\in E} w(e)|e|},\]
where $w$ ranges over weight functions on $E$ with finite support.

Our goal is to show $\tau(\EE) = \gap(\EE)$.
Since we know $\gap(\EE) \le \tau(\EE)$, it suffices to show
$\tau(\EE) \le \gap(\EE)$.

Define $\E$ to contain those edges in $\EE$ with rational endpoints.
(In fact, any countable {\em dense} set will do in place of the rationals.)
Fix any enumeration of the edges in $\E$.
Let set $\E_n$ contain the first $n$ edges in the enumeration.
It suffices to show the following equality:
\begin{equation}\label{twostep}
  \sup_n \tau(\E_n) = \tau(\EE),
\end{equation}
because by Theorem~\ref{finite_case} we know $\tau(\E_n) = \gap(\E_n)$,
and clearly $\gap(\E_n) \le \gap(\EE)$,
so combining with~(\ref{twostep}) proves the theorem via
\[
  \tau(\EE) = \sup_n \tau(\E_n) = \sup_n \gap(\E_n) \le \gap(\EE).
\]
Thus, to prove Theorem~\ref{infinite_case},
we need only prove~(\ref{twostep}).
In the remainder of the section we prove it as follows:
We first show we can extend any sequence of cutting
schemes, one for each $\E_n$,
to a single good cutting scheme for their union $\E$
(edges with rational endpoints).
This shows $\sup_n \tau(\E_n) = \tau(\E)$.
We then show that we can extend the cutting scheme
on $\E$ to a cutting scheme on $\EE$ (all edges).
This shows $\tau(\E) = \tau(\EE)$.

One measure-theoretic issue that we must first address is
how we formally define a probability distribution $P$
on our infinite sets $\EE$ and $\E$.
For this we use Kolmogorov's Existence Theorem
\cite[ch.~7]{Bill}.

Before we explain, we introduce some terminology.
Note that any cutting scheme $P$ of a set $F \subseteq \EE$
induces a cutting scheme $P_E$ on each subset $E\subseteq F$
by restriction.  We say that  any $P$ from which $P_E$ can
be so obtained {\em is consistent with} $P_E$.

Kolmogorov's Existence Theorem implies the following.
Fix any $F\subseteq \EE$.
Consider a family
of cutting schemes $\langle P_E : E \subseteq F, E \mbox{ finite}\rangle$
of the finite subsets of $\EE$.
If this family is {\em consistent}, meaning that
whenever two cutting schemes $P_E$ and $P_{E'}$ in the family
satisfy $E\subseteq E'$, the second scheme is consistent with the first one,
then there exists a single cutting scheme $P$ of $F$,
with $P$ consistent with each $P_E$ in the family.

In this section, to describe any cutting
scheme on $\EE$ (or $\E$), we specify
the consistent family of cutting schemes it induces on the finite subsets.
The following lemma gives a useful condition for the existence
of such a family.  We will use it twice.
\begin{lemma}\label{converge}
  Let $F$ be any subset of $\EE$.
  Suppose there exists a sequence of cutting schemes
  $\langle Q^{(1)},Q^{(2)},\ldots\rangle$ that {\em converges}
  in the following sense: for each finite $E\subseteq F$
  and each $k$-way cut $C$ of $E$, the limit
\begin{equation} \label{prob-limit}   \lim_{n\rightarrow\infty}
  Q^{(n)}_E(C)
\end{equation}
  is well defined.  Define $P_E$ to be the cutting scheme of $E$ that, for
  each $k$-way cut $C$ of $E$, chooses $C$ with probability
  in~(\ref{prob-limit}), i.e. $P_E(C) := \lim_{n\rightarrow\infty}
  Q^{(n)}_E(C)$.
  Then
  $\langle P_E : E \subseteq F, E \mbox{ finite}\rangle$
  is a consistent family of cutting schemes,
  so by Kolmogorov's Existence Theorem
  there exists a cutting scheme $P$ of $F$ consistent with each $P_E$.
\end{lemma}
\begin{proof}
  We need to verify the following:
  \begin{enumerate}
  \item Each $P_E$ is a cutting scheme:
    $\sum_C P_E(C) = 1$ with each $P_E(C) \ge 0$,
    where $C$ ranges over all cuts of \newmt{$E$}{$C$}{}.
  \item For $P_E$ and $P_{E'}$ with $E\subseteq E'$ both finite,
    $P_{E'}$ is consistent with $P_E$.    That is, for any cut $C$ of
    $E$,
    \[P_E(C) = \sum_{C'} P_{E'}(C'),\]
    where $C'$ ranges over all $k$-way cuts of $E'$ such that $C' \cap E = C$.
  \end{enumerate}
  In each case, the desired property holds for the cutting scheme
  induced on $E$ (and/or $E'$) by $Q^{(m)}$ for large enough $m$.
  For example, to verify the second property, use
  \begin{eqnarray*}
    P_E(C) - \sum_{C'} P_{E'}(C')
    &=&
    \lim_{n} Q^{(n)}_E(C) \,-\, \sum_{C'}  \lim_{n} Q^{(n)}_{E'}(C') 
    \\ &=&
    \lim_{n} \Big[Q^{(n)}_E(C) - \sum_{C'}  Q^{(n)}_{E'}(C')\Big]
    \\ &=& 0.
  \end{eqnarray*}
  The first equality is by definition of $P_E$,
  the second is because the \newmt{finite}{}{} sum of the limits is the 
  limit of
  the sums,
  and the last is because, for $n$ large enough, the term inside the limit
  is well defined and necessarily 0 (simply because $Q^{(n)}$
  is a cutting scheme, so the induced cutting schemes 
  $Q^{(n)}_E$ and   $Q^{(n)}_{E'}$ 
  are necessarily consistent).
  The first property follows similarly.
\end{proof}

We need one last ``utility'' lemma.
It will help us construct a sequence
of cutting schemes $\langle Q^{(i)}\rangle$ that converges
in the sense needed for Lemma~\ref{converge}.

\begin{lemma}\label{infinite}
  Consider a countable collection $\mathcal C$
  of countable sequences of real numbers, where each sequence 
  $q = \langle q^{(1)},q^{(2)},\ldots \rangle$ in $\mathcal C$
  lies in some finite interval $I_s$.
  
  Then there exists an infinite index set $\mathcal I \subseteq \{1,2,\ldots\}$
  such that for each sequence $q \in \mathcal C$, the limit
  \[\lim_{n\in \mathcal I, n\rightarrow\infty} q^{(n)}\]
  is well defined.
\end{lemma}
\begin{proof}
  The proof is a ``dovetailing'' variation of the standard
  proof that any sequence in a compact set contains an
  infinite convergent subsequence.
  
  Order the sequences arbitrarily and consider the infinite
   matrix whose $j$th row is the $j$th sequence.
  Associate with each sequence $q$ an interval $I'_q$, initially $I_q$.
  Visit the sequences in the standard dovetailing order;
  that is, visit the $j$th sequence for each $j=1,1,2,1,2,3,1,2,3,4,\ldots$,
  so that each sequence is visited infinitely often.
  
  While visiting a sequence $q$, narrow the associated
  interval to either its upper or lower half, and then delete
  from the matrix all the columns $i$ such that $q^{(i)}$ is no longer in the
  associated interval.  Further, make the choice of upper or
  lower half so that infinitely many columns remain undeleted.
  (This is possible because each of the infinitely many
  $q^{(i)}$'s lying in a column that was previously not deleted
  lies in one of the two halves.)

  To complete the construction, define $i(n)$ to be the
  smallest index larger than $n$ of any column that remains undeleted after the $n$th step
  of the construction, and take $\mathcal I=\{i(n) : n=1,2,\ldots\}$.
  It is easy to verify that for each $q\in\mathcal C$, the subsequence
  $\langle q^{(i)} : i\in \mathcal I\rangle$
  converges.
\end{proof}

Now we can begin our two-step proof
that $\tau(\EE) = \tau(\E) =\sup_n \tau(\E_n)$.
First we show that if each $\E_n$ has a good cutting scheme,
then there is an equally good cutting scheme of $\E$.
\begin{lemma}\label{stepone}
  There exists a cutting scheme $P$ of $\E$
  with maximum density $\sup_n \tau(\E_n)$.
  Thus, $\tau(\E) =\sup_n \tau(\E_n)$
\end{lemma}
\begin{proof}
  Let $\tau_{\sup} = \sup_n \tau(\E_n)$.
  Say a sequence of cutting schemes $\langle P^{(n)}\rangle$ is {\em good} if
  each $P^{(n)}$ is a cutting scheme of $\E_n$ and has maximum
  density at most $\tau_{\sup}$ with respect to $\E_n$.
  By the definition of $\tau_{\sup}$, there exists a 
  good sequence of cutting schemes $\langle Q^{(n)} : n=1,2,\ldots \rangle$.

  Consider each pair $(E,C)$ where $E$ is a finite subset of
  $\E$ and $C$ is a cut of $E$.
  Since there are only countably many such pairs,
  Lemma~\ref{infinite} implies that there exists an infinite
  index set $\mathcal I\subseteq\{1,2,\ldots\}$
  such that, for each pair \newmt{$(E,C)$}{}{},
  \[P_E(C) := \lim_{n\in \mathcal I, n\rightarrow\infty} Q^{(n)}_E(C)\]
  \newmt{}{(where $n\rightarrow\infty$)}{put it under lim above}
  is well-defined.
  By Lemma~\ref{converge},
  each $P_E$ is a cutting scheme
  and there exists a cutting scheme $P$ of $\E$
  that is consistent with each $P_E$.

  It remains to verify that $P$ has maximum density
  $\tau_{\sup}$ with respect to $\E$.  Observe that, for any edge $e \in \E$ and any cutting scheme $Q$,
$Q_{\set{e}}$ is a cutting scheme that chooses among at most two 
cuts, the set $\set{e}$ itself and (possibly) the empty set, and that
the probability that $Q_{\set{e}}$ chooses the set $\set{e}$ is the
probability that $Q$ chooses some cut that contains $e$.

For any edge $e \in \E$,
\begin{eqnarray*}
\Pr[P \mbox{ cuts } e] & = & \Pr[P_{\set{e}} \mbox{ chooses the cut } \set{e}]\\
& = & \lim_{n\in \mathcal I, n\rightarrow\infty}
Q^{(n)}_{\set{e}}(\set{e})\\
& = & \lim_{n\in \mathcal I, n\rightarrow\infty}
\Pr[Q^{(n)} \mbox{ cuts } e]\\
& \leq & \lim_{n\in \mathcal I, n\rightarrow\infty}
|e|\tau_{\sup}
\end{eqnarray*}
 Hence, for any edge $e$,
the probability that $P$ cuts $e$ is at most $|e|\tau_{\sup}$.
\end{proof}

\begin{lemma}
  Let $P$ be the cutting scheme of $\E$
  of maximum density $\tau(\E)$ (shown to exist in Lemma~\ref{stepone}).
  There is a cutting scheme $\overline{P}$ of $\EE$ that has maximum density $\tau(\E)$ with respect to $\EE$.
  Thus $\tau(\EE) =\tau(\E)$.
\end{lemma}
\begin{proof}
  For each point $p$ in the simplex, fix a sequence of rational points
  $\langle p^{(1)},p^{(2)},\ldots\rangle$ converging to $p$.
  For any set $F\subseteq \EE$,
  let $F^{(n)}$ denote $\{(p^{(n)}, q^{(n)}) : (p,q)\in F\}$.
  
  Define a sequence of cutting schemes
  $\langle P^{(n)} : n=1,2,\ldots\rangle$ of $\EE$
  by $P^{(n)}(C) = P(C^{(n)})$.
  That is, $P^{(n)}$ maps each real point to its
  $n$th rational approximation, then cuts using
  a random cut from $P$.
  We claim that for each finite $E\subseteq \EE$
  and each cut $C$ of $E$, the limit
  \[
  \lim_{n\rightarrow\infty} P^{(n)}_E(C)
  \,=\,
  \lim_{n\rightarrow\infty} P_{E^{(n)}}(C^{(n)})
  \] is well defined.

  To show the claim, it is enough to show that for all $\epsilon>0$,
  $|P_{E^{(n)}}(C^{(n)}) - P_{E^{(m)}}(C^{(m)})| \le \epsilon$
  for all large enough $n$ and $m$.
  Let $E' = E^{(n)} \cup E^{(m)}$
  and $C' = C^{(n)} \cup C^{(m)}$.
  By consistency,
  \[P_{E'}(C') \,\le\, P_{E^{(n)}}(C^{(n)})
  \,\le\, P_{E'}(C') + \sum_{D} P_{E'}(D),\]
  where $D$ ranges over all the cuts of $E'$
  other than $C'$ that are consistent with $C$.
  Each of the cuts that $D$ takes on cuts
  at least one of the segments $(p^{(n)}, p^{(m)})$
  where $p$ ranges over the endpoints of the edges in $E$,
  so the corresponding term in the sum
  is at most $\tau(\E)|p^{(n)}- p^{(m)}|$.
  Since there are finitely many terms and $\tau(\E)$ is finite
  and $p^{(n)}\rightarrow p$, it follows that
  the sum on the right is arbitrarily small for large enough $n$ and $m$.
  Thus $|P_{E^{(n)}}(C^{(n)}) - P_{E'}(C')|$
  (and likewise $|P_{E^{(m)}}(C^{(m)}) - P_{E'}(C')|$)
  tends to zero for large $m$ and $n$.
  Thus so does $|P_{E^{(n)}}(C^{(n)}) - P_{E^{(m)}}(C^{(m)})|$.  This proves the claim.

Define $\overline{P}_E(C) := \lim_{n\rightarrow\infty} P^{(n)}_E(C)$.
By Lemma 7.4, there is a cutting scheme $\overline{P}$ of $\EE$ that is
consistent with each $\overline{P}_E$.  The argument used at the end
of the proof of Lemma~\ref{stepone} shows that, for any edge $e \in
\EE$, 
\begin{eqnarray*}
\Pr[P \mbox{ cuts }e] & = & \lim_{n\rightarrow\infty}
P^{(n)}_{\set{e}}(C)\\
& = & \lim_{n\rightarrow\infty} P_{{\set{e}}^{(n)}}(C^{(n)})\\
& = & \lim_{n\rightarrow\infty} \Pr[P \mbox{ cuts } e^{(n)}]\\
& \leq & \lim_{n\rightarrow\infty} \tau(P) |e^{(n)}|\\
& = & \tau(P) \lim_{n\rightarrow\infty}  |e^{(n)}|\\
& = & \tau(P) |e|
\end{eqnarray*}
Thus the maximum density of $\overline{P}$ is at most $\tau(P)$.
\end{proof}

\end{document}

%% file: threefig.tex
\newcommand{\AX}{150}\newcommand{\AY}{210}
\newcommand{\Alabel}{1}
\newcommand{\BX}{100}\newcommand{\BY}{140}
\newcommand{\Blabel}{a}
\newcommand{\CX}{200}\newcommand{\CY}{140}
\newcommand{\Clabel}{b}
\newcommand{\DX}{50}\newcommand{\DY}{70}
\newcommand{\EX}{250}\newcommand{\EY}{70}
\newcommand{\FX}{0}\newcommand{\FY}{0}
\newcommand{\Flabel}{2}
\newcommand{\GX}{100}\newcommand{\GY}{0}
\newcommand{\Glabel}{c}
\newcommand{\HX}{200}\newcommand{\HY}{0}
\newcommand{\Hlabel}{d}
\newcommand{\IX}{300}\newcommand{\IY}{0}
\newcommand{\Ilabel}{3}

\newcommand{\JX}{50}\newcommand{\JY}{70}
\newcommand{\Jlabel}{e}
\newcommand{\KX}{250}\newcommand{\KY}{70}
\newcommand{\Klabel}{f}

\newcommand{\PX}{100}\newcommand{\PY}{90}
\newcommand{\Plabel}{r}
\newcommand{\QX}{80}\newcommand{\QY}{118}
\newcommand{\Qlabel}{q}
\newcommand{\RX}{60}\newcommand{\RY}{90}
\newcommand{\Rlabel}{q'}
\newcommand{\SX}{170}\newcommand{\SY}{188}
\newcommand{\Slabel}{s}
\newcommand{\TX}{240}\newcommand{\TY}{90}
\newcommand{\Tlabel}{s'}
\newcommand{\UX}{30}\newcommand{\UY}{-8}
\newcommand{\Ulabel}{t}
\newcommand{\VX}{170}\newcommand{\VY}{-8}
\newcommand{\Vlabel}{t'}

\newcommand{\WX}{15}\newcommand{\WY}{36}
\newcommand{\Wlabel}{w}
\newcommand{\XX}{45}\newcommand{\XY}{-8}
\newcommand{\Xlabel}{x}
\newcommand{\YX}{255}\newcommand{\YY}{-8}
\newcommand{\Ylabel}{y}
\newcommand{\ZX}{285}\newcommand{\ZY}{36}
\newcommand{\Zlabel}{z}

\newcommand{\doTriangle}{%
  \drawline(\IX,\IY)(\AX,\AY)(\FX,\FY)(\IX,\IY)  % large outer triangle
}
\newcommand{\doTriangleLabels}{%
  \put(\AX,\AY){\makebox(0,0)[br]{$\Alabel$}}
  \put(\FX,\FY){\makebox(0,0)[tr]{$\Flabel$}}
  \put(\IX,\IY){\makebox(0,0)[tl]{$\Ilabel$}}
}
\newcommand{\doBallCut}{%
  \drawline(\PX,\PY)(\RX,\RY)
  \drawline(\PX,\PY)(\TX,\TY)
  \drawline(\PX,\PY)(\VX,\VY)
  \dottedline{5}(\QX,\QY)(\PX,\PY)
  \dottedline{5}(\UX,\UY)(\PX,\PY)
  \dottedline{5}(\SX,\SY)(\PX,\PY)
  \put(\PX,\PY){\circle*{10}}
}
\newcommand{\doBallCutLabels}{%
  \put(\PX,\PY){\makebox(0,0)[t]{\shortstack{~\\~\\$\Plabel$}}}
  \put(\QX,\QY){\makebox(0,0)[br]{$\Qlabel$}}
  \put(\RX,\RY){\makebox(0,0)[r]{$\Rlabel$}}
  \put(\SX,\SY){\makebox(0,0)[bl]{$\Slabel$}}
  \put(\TX,\TY){\makebox(0,0)[l]{$\Tlabel$}}
  \put(\UX,\UY){\makebox(0,0)[tr]{$\Ulabel$}}
  \put(\VX,\VY){\makebox(0,0)[tl]{$\Vlabel$}}
  }
\newcommand{\doUnif}{%
  \put(\PX,\PY){\circle*{10}}
  \dottedline{3}(\GX,\GY)(\BX,\BY)
  \dottedline{3}(\CX,\CY)(\HX,\HY)
  }
\newcommand{\doUnifLabels}{%
  \put(\BX,\BY){\makebox(0,0)[br]{$\Blabel$}}
  \put(\CX,\CY){\makebox(0,0)[bl]{$\Clabel$}}
  \put(\GX,\GY){\makebox(0,0)[t]{$\Glabel$}}
  \put(\HX,\HY){\makebox(0,0)[t]{$\Hlabel$}}
  \put(\PX,\PY){\makebox(0,0)[l]{~$\Plabel$}}
  \put(\JX,\JY){\makebox(0,0)[br]{$\Jlabel$}}
  \put(\KX,\KY){\makebox(0,0)[bl]{$\Klabel$}}
}
\newcommand{\doCorners}{%
  \dottedline{4}(\DX,\DY)(\GX,\GY) % lower left corner
  \dottedline{4}(\BX,\BY)(\CX,\CY) % upper corner
  \dottedline{4}(\HX,\HY)(\EX,\EY) % lower right corner
  }
\newcommand{\doCornerCut}{%
  \drawline(\WX,\WY)(\XX,\XY)
  \drawline(\YX,\YY)(\ZX,\ZY)
  }
\newcommand{\doCornerCutLabels}{%
  \put(\WX,\WY){\makebox(0,0)[br]{$\Wlabel$}}
  \put(\XX,\XY){\makebox(0,0)[tl]{$\Xlabel$}}
  \put(\YX,\YY){\makebox(0,0)[tr]{$\Ylabel$}}
  \put(\ZX,\ZY){\makebox(0,0)[bl]{$\Zlabel$}}
  }